\documentclass[11pt]{article}
\usepackage{epsf}
\usepackage{epsfig}
\newcommand{\braket}[2]{\langle #1 | #2 \rangle}
\newcommand{\ket}[1]{\left | \, #1 \right \rangle}

\newcommand{\bra}[1]{\left \langle #1 \, \right |}
\newcommand{\proj}[1]{\ket{#1}\!\!\bra{#1}}
\newcommand{\Tr}{\mbox{\rm \/Tr }}

\newtheorem{conjecture}{Conjecture}
\begin{document}
\begin{center}
{{\LARGE The Adaptive Classical Capacity of a Quantum Channel,\large \\ or\\ 
\Large Information Capacities of Three Symmetric Pure 
States \\ in Three Dimensions\\
\large Peter W. Shor\\
}AT\&T Labs---Research\\
Florham Park, NJ 07932
}
\end{center}

{\abstract{
We investigate the capacity of three symmetric quantum states
in three real dimensions to carry classical information.  Several
such capacities have already been defined, depending on what operations
are allowed in the sending and receiving protocols.  These include
the $C_{1,1}$ capacity, which is the capacity achievable if separate 
measurements must be used for each of the received states, and the 
$C_{1,\infty}$ capacity, which is the capacity achievable if joint 
measurements are allowed on the tensor product
of all the received states.
We discover a new classical information capacity of quantum channels, 
the {\em adaptive} capacity $C_{1,A}$, which lies strictly between 
the $C_{1,1}$ and the $C_{1,\infty}$ capacities.  
  The adaptive capacity allows what is known as the LOCC (local
operations and classical communication) model of quantum
operations for decoding the channel outputs.  This model requires each 
of the signals to be measured by a separate apparatus, but allows the 
quantum states of these signals to be measured in stages, with the
first stage partially reducing their quantum states, and where
measurements in subsequent stages 
may depend on the results of a classical computation taking as input the
outcomes of the first
round of measurements.
We also show that even in three dimensions, with the information carried
by an ensemble containing three pure states,
achieving the $C_{1,1}$ capacity may require a POVM with 
six outcomes.
}}

\section{Introduction}
\label{sec-intro}

For classical channels, Shannon's theorem \cite{Shannon48} gives 
the information-carrying capacity of a channel.  When one tries
to generalize this to quantum channels, there are several ways to
formulate the problem which have given rise to several different
capacities.  In this paper, we consider the capacity of quantum
channels to carry classical information, with various restrictions
on how the channel may be used.  Several such capacities have 
already been defined for quantum channels.  In particular, the $C_{1,1}$
capacity, where only tensor product inputs and tensor product measurements
are allowed \cite{fuchs-thesis,Levitin-ai,Levitin-conj,Osaki-ai}, 
and the $C_{1,\infty}$ capacity, where tensor product
inputs and joint measurements are allowed 
\cite{Hol-cap,SW,Holevo-codingtheorems}, 
have both been studied
extensively.  We
will be investigating these capacities
in connection with a specific example;
namely, we analyze how these capacities behave on a symmetric set of three 
quantum states in three dimensions which we call the lifted trine states.  
A quantum 
channel of the type that Holevo \cite{Holevo-codingtheorems} classifies as 
c-q (classical-quantum) can be constructed from these states 
by allowing the sender to choose one of these three pure
states, which is then conveyed to the receiver.
This channel is simple enough that we can analyze the behavior of 
various capacities for it, but it is also complicated enough to 
exhibit interesting behaviors which have not been observed before.  
In particular, we define a new, natural, classical capacity for a 
quantum channel, the $C_{1,A}$ capacity, which we also call 
the {\em adaptive one-shot capacity}, and show that it is strictly 
between the $C_{1,1}$ capacity (also called the one-shot quantum 
capacity) and the $C_{1,\infty}$ capacity
(also called the Holevo capacity).

The three states we consider, the 
lifted trine states, are obtained by starting with the two-dimensional
quantum trine states, $(1,0)$, $(-1/2,\sqrt{3}/2)$, $(-1/2, -\sqrt{3}/2)$
introduced by Holevo \cite{Holevo-trines} and later studied by Peres 
and Wootters \cite{Peres-Wootters}.
We add a third dimension to the Hilbert space of the trine states,
and lift all of the trine
states out the plane into this dimension by
an angle of $\arcsin \sqrt{\alpha}$, so the states become 
$(\sqrt{1-\alpha}, 0, \sqrt{\alpha})$, and so forth.
We will be dealing with small $\alpha$ (roughly, $\alpha < 0.1$),
so that they are close to being
planar.   This is one of the interesting regimes.  When 
the trine states are lifted further out of the plane, they behave
in less interesting ways until they are close to being
vertical; then they start being interesting again, but we
will not investigate this second regime.

To put this channel into the formulation of completely positive 
trace-preserving operators,
we let the sender
start with a quantum state in a 
three-dimensional input state space, measure this state using a von
Neumann measurement with three outcomes, 
and send one of the lifted trines $T_0$, $T_1$ or $T_2$,
depending on the outcome of this measurement.  This process turns any
quantum state into a probability distribution over $T_0$, $T_1$ and~$T_2$.

The first section of the paper deals
with the accessible information for the lifted trine states when the
probability of all three states is equal.  The accessible information of
an ensemble is the maximum mutual information obtainable
between the input states 
of the ensemble and the outcomes of a POVM (positive operator 
valued measure) measurement
on these states.  The substance of this section has already appeared,
in \cite{shor-paper1}.  Combined with Appendix C, this shows that the
number of projectors required to achieve the $C_{1,1}$ capacity for
the ensemble of lifted trines can be as large as 6, the maximum possible
by the real version of Davies' theorem.  
The second section deals with the $C_{1,1}$ channel capacity (or the
one-shot capacity), which is 
the maximum of the accessible information over all probability distributions 
on the trine states.  This has often been called the $C_1$ capacity
because it is the classical capacity obtained
when you are only allowed to process (i.e., encode/measure)
one signal at a time.  We call it $C_{1,1}$ to emphasize that you are 
only allowed to input tensor product states (the first 1), and only 
allowed to make quantum measurements on one signal at a time (the second 1).
The third section deals with the new capacity $C_{1,A}$, the
``adaptive one-shot capacity.''
This is 
the capacity for sending classical information attainable
if you are allowed to send codewords composed of tensor products
of lifted trine states, are not allowed to make joint measurements 
involving more than one trine state, but are allowed to
make a measurement on one signal which only partially reduces the quantum 
state, use the outcome of this measurement to determine which 
measurement to make on a different signal, return to refine the measurement 
on the first signal, and so forth.  
In Section~\ref{sec-exp-close}, we give an upper bound on the 
$C_{1,1}$ capacity of the lifted trine states, letting us show 
that for the lifted trine states 
with sufficiently small $\alpha$, this adaptive capacity is strictly larger
than the $C_{1,1}$ channel capacity.  
In section \ref{sec-upperbound}, we show
and that for two pure non-orthogonal states, $C_{1,A}$
is equal to $C_{1,1}$, and thus 
strictly less than the Holevo capacity $C_{1,\infty}$.
These two results show
show that $C_{1,A}$ is different from
previously defined capacities for quantum channels.
To obtain a capacity larger than $C_{1,1}$, it is necessary
to make measurements that only partially reduce the state of some of the
signals, and 
then later return to refine the measurement on these signals depending
on the results of intervening measurement.  
In Section \ref{sec-discussion}, we show
if you use ``sequential measurement'', 
i.e., only
measure one signal at a time, and never return to a previously measured signal, 
it is impossible to achieve a capacity larger than $C_{1,1}$.  

We take the lifted trine states 
to be:
\begin{eqnarray}
T_0(\alpha) &=& (\sqrt{1-\alpha}, 0, \sqrt{\alpha})
\nonumber
\\
T_1(\alpha) &=& (-{\textstyle{\frac{1}{2}}}\sqrt{1-\alpha}, 
{\textstyle{\frac{\sqrt{3}}{2}}}\sqrt{1-\alpha}, \sqrt{\alpha})\\
T_2(\alpha) &=& (-{\textstyle{\frac{1}{2}}}\sqrt{1-\alpha}, 
- {\textstyle{\frac{\sqrt{3}}{2}}}\sqrt{1-\alpha}, 
\sqrt{\alpha})
\nonumber
\end{eqnarray}
When it is clear what $\alpha$ is, we may drop it from the notation
and use $T_0$, $T_1$, or~$T_2$.

\section{The Accessible Information}
\label{sec-accessible}

In this section, we find the accessible information for the ensemble
of lifted trine states, given equal probabilities.  This is defined
as the maximal
mutual information between the trine states (with probabilities $\frac{1}{3}$
each) and the elements of a POVM measuring these states.
Because the trine states are vectors over the reals, it follows from the
generalization of Davies' theorem to real states 
(see, e.g., \cite{Davies-gen}) that there is an optimal POVM
with at most six elements, all the components of which are real.  
The lifted trine states are three-fold symmetric, so by symmetrizing we
can assume that the optimal POVM is three-fold symmetric (possibly at
the cost of introducing extra POVM elements).  Also, 
the optimal POVM can be taken to have one-dimensional elements $E$, 
so the elements
can be described as vectors $\ket{v_i}$ where $E_i = \ket{v_i} \bra{v_i}$.  
This means that there is an 
optimal POVM whose vectors come in triples of the form: 
$\sqrt{p} P_0(\phi,\theta)$,
$\sqrt{p} P_1(\phi,\theta)$,
$\sqrt{p} P_2(\phi,\theta)$,
where $p$ is a scalar probability and 
\begin{eqnarray}
\nonumber
P_0(\phi, \theta) &=& (\cos\phi\cos\theta, 
\cos\phi\sin\theta, \sin\phi)
\label{definePs}
\\
P_1(\phi, \theta) &=&  
(\cos\phi\cos(\theta + 2 \pi/3), 
\cos\phi\sin(\theta + 2 \pi/3), \sin\phi)
\\
P_2(\phi, \theta) &=& 
(\cos\phi\cos(\theta - 2 \pi/3), 
\cos\phi\sin(\theta - 2 \pi/3), \sin\phi).
\nonumber
\end{eqnarray}

The optimal POVM may have several such triples, which we label
$\sqrt{p_1}\, P_b(\phi_1, \theta_1)$, 
$\sqrt{p_2}\, P_b(\phi_2, \theta_2)$, $\ldots$,
$\sqrt{p_m}\, P_b(\phi_m, \theta_m)$.
It is easily seen that
the conditions for this set of vectors to be a POVM are that
\begin{equation}
    \sum_{i=1}^m   p_i \sin^2(\phi_i) = 1/3                    
{\mathrm{\quad and\quad }}
    \sum_{i=1}^m   p_i  = 1.
\label{constraint1}
\end{equation}

One way to compute the accessible information $I_A$ is to
break the formula for accessible information 
into pieces so as to keep track of the amount of information
contributed to it by each triple.
That is, $I_A$ will be the weighted average (weighted by $p_i$) of the 
contribution $I(\phi,\theta)$
from each $(\phi, \theta)$.  To see this, recall that $I_A$ is the
mutual information between the input and the output, and that
this can be expressed
as the entropy of the input less the entropy of the input given the output,
$H(X_\mathrm{in}) -  H(X_\mathrm{in}|X_\mathrm{out})$.
The term $H(X_\mathrm{in}|X_\mathrm{out})$ naturally decomposes into terms
corresponding to the various POVM outcomes, and there are several ways of
assigning the entropy of the input
$H(X_\mathrm{in})$ 
to these POVM elements in order to complete this 
decomposition.  This is how I first arrived at the formula for $I_A$.
I briefly sketch this analysis,
and then go into detail in a second analysis.  This second 
analysis is superior in that it
explains the form of the answer obtained, but it is not clear how one
could discover the second analysis without first knowing the result.

For each $\phi$, and each $\alpha$, there is a $\theta$ that optimizes
$I(\phi,\theta)$.  
This $\theta$ starts out
at $\pi/6$ for $\phi=0$, decreases until it hits 0 at some value of $\phi$
(which depends on $\alpha$), 
and stays at $0$ until $\phi$ reaches its maximum value of $\pi/2$.  
For a fixed $\alpha$, by finding (numerically)
the optimal value of $\theta$ for each $\phi$ and using it to obtain
the contribution to $I_A$ attributable to that $\phi$, we get a curve 
giving the optimal contribution to $I_A$ for each $\phi$.  
If this curve is plotted,
with the $x$-value being $\sin^2 \phi $ and the 
$y$-value being the contribution to $I_A$,
an optimal POVM can be obtained by finding the set of points on this curve
whose average $x$-value is $1/3$ (from Eq.~\ref{constraint1}), and whose
average $y$-value is as large as possible.
A convexity
argument shows that we only need at most two points from the curve
to obtain this
optimum; we will need one or two points depending on whether the
relevant part of the curve is concave or convex.  
For small $\alpha$, it can be seen numerically
that the relevant piece of the
curve is convex, and we need two $\phi$'s to
achieve the maximum.  
One of the $(\phi,\theta)$ pairs is $(0,\pi/6)$, 
and the other is $(\phi_{\alpha},0)$ for some  
$\phi_\alpha > \arcsin(1/\sqrt{3})$.  
The formula for this
$\phi_\alpha$ will be derived later.
Each of these $\phi$'s corresponds to a triple of POVM elements, giving
six elements for the optimal POVM.   

The analysis in the remainder of this section gives a different way
of describing this six-outcome POVM.  This analysis unifies the 
measurements for different $\alpha$, and also
introduces some of the methods that will appear again in 
Section~\ref{sec-adaptive}.
Consider the following measurement protocol.
For small $\alpha$ ($\alpha < \gamma_1$ for some constant~$\gamma_1$), 
we first 
take the trine $T_b(\alpha)$ and make a partial
measurement which either
projects it onto the $x,y$ plane or lifts it further out of this
plane so that it becomes
the trine $T_b(\gamma_1)$.
(Here $\gamma_1$ is
independent of $\alpha$.)  If the measurement projects the trine into the
$x,y$ plane, we make a second measurement using the POVM having 
outcome vectors $\sqrt{2/3}(0,1)$  and
$\sqrt{2/3} (\pm\sqrt{3}/2, -1/2)$.  This is the optimal POVM for trines
in the $x,y$-plane.  
If the measurement lifts the trine further out of the $x,y$ plane, we use 
the von Neumann measurement that projects onto the basis consisting of
$(\sqrt{2/3}, 0, \sqrt{1/3})$,
$(-\sqrt{1/6}, \pm \sqrt{1/2}, \sqrt{1/3})$.  
If $\alpha$ 
is larger than $\gamma_1$ (but smaller than $8/9$), 
we skip the first partial measurement, and just
use the above von Neumann 
measurement.   Here, $\gamma_1$
is obtained by numerically solving a fairly complicated equation;
we suspect that no closed form expression for it exists.  
The value of $\gamma_1$ is 0.061367, which is $\sin^2 \phi$ for $\phi \approx
0.25033$ radians ($14.343^\circ$).

We now give more details on this decomposition of
the POVM into a two-step
process.  We first apply a partial measurement which does not extract
all of the quantum information, i.e., it leaves a quantum residual state
that is not completely determined by the measurement outcome.
Formally, we apply one of a set
of matrices $A_i$ satisfying $\sum_i A_i^\dagger  A_i= I$.  If we start
with a pure state $\ket{v}$, we observe the $i$'th outcome with probability
$\bra{v} A_i^\dagger A_i \ket{v}$, and in this case the state $\ket{v}$ 
is taken to the state $A_i \ket{v}$.  We choose as the
$A_i$'s the matrices $\sqrt{p_i}\, M(\phi_i)$ where
\begin{equation}
M(\phi) = 
\left(
\begin{array}{ccc}
\sqrt{\frac{3}{2}} \cos \phi  & 0 & 0 \\
0 &  \sqrt{\frac{3}{2}} \cos \phi & 0 \\
0 & 0 & \sqrt{3} \sin \phi 
\end{array}
\right)
\end{equation}
The $\sqrt{p_i}\,M(\phi_i)$ form a valid partial measurement
if and only if 
\[
\sum_i p_i \sin^2(\phi_i) = 1/3 \mathrm{\ \ \ and \ \ }\sum_i p_i = 1.
\]
By first applying the above $\sqrt{p_i}\,M(\phi_i)$, and then applying 
the von Neumann 
measurement with the three basis vectors
\begin{eqnarray}
\nonumber
V_0(\theta) &= & \textstyle \Big(\sqrt{\frac{2}{3}}\cos\theta, 
\sqrt{\frac{2}{3}}\sin\theta, \frac{1}{\sqrt{3}}\Big) \\
\label{defineV}
V_1(\theta) &=& \textstyle \Big(\sqrt{\frac{2}{3}}\cos(\theta+2 \pi/3 ), 
\sqrt{\frac{2}{3}}\sin(\theta+2 \pi/3), \frac{1}{\sqrt{3}}\Big) \\
V_2(\theta) &=& \textstyle \Big(\sqrt{\frac{2}{3}}\cos(\theta-2 \pi/3 ), 
\sqrt{\frac{2}{3}}\sin(\theta-2 \pi/3), \frac{1}{\sqrt{3}}\Big) 
\nonumber
\end{eqnarray}
we obtain the POVM given by
the vectors $\sqrt{p_i} \, P_b(\theta_i, \phi_i)$ of Eq.~(\ref{definePs});  
checking this is simply a matter of verifying that 
$V_b(\theta) M(\phi) = P_b(\theta,\phi)$.  
Now, after applying $\sqrt{p_i}\, M(\phi_i)$ to the trine $T_0(\alpha)$, 
we get the vector
\begin{equation}
\big(\sqrt{3/2}\sqrt{1-\alpha}\sqrt{p_i} \cos \phi_i , 0, 
\sqrt{3} \sqrt{\alpha} \sqrt{p_i} \sin \phi_i  \big).
\end{equation}
This is just the state
$\sqrt{p_i'}\, T_0(\alpha_i')$ where $T_0(\alpha_i')$ is the trine state
with
\begin{equation}
\alpha_i' = \frac{\alpha \sin^2 \phi_i }{\alpha \sin^2 \phi_i
+ \frac{1}{2} (1-\alpha) \cos^2\phi_i},
\label{alpha-prime}
\end{equation}
and where
\begin{equation}
p_i'  = 3 p_i \left[ \alpha \sin^2  \phi + {\textstyle{\frac{1}{2}}}(1-\alpha) 
\cos^2 \phi \right]
\label{p-prime}
\end{equation}
is the probability that we observe this trine state, given that we started
with $T_0(\alpha)$.  Similar
formulae hold for the trine states $T_1$ and $T_2$.  
We compute that
\begin{equation}
\sum_i p_i' \alpha_i' = \sum_i 3 p_i \alpha \sin^2 (\phi_i) = \alpha.
\end{equation}
The first stage of this process, 
the partial measurement which applies the matrices
$\sqrt{p_i}\, M(\phi_i)$, reveals
no information about which of $T_0$, $T_1$, $T_2$ 
we started with.  Thus, by the
chain rule for classical Shannon information \cite{Cover}, 
the accessible information obtained by our two-stage
measurement is just the weighted average (the weights being
$p'_i$) of the maximum over $\theta$ of the
Shannon mutual information $I_{\alpha_i'}(\theta)$ between 
the outcome of the von Neumann 
measurement $V(\theta)$ and the trine $T(\alpha_i')$.  
By convexity, it suffices to use 
only two values of $\alpha_i'$ to obtain this maximum.  In fact, the optimum
is obtained using either
one or two values of $\alpha_i'$ depending on whether the function
\[
I_{\alpha'} = \max_\theta I_{\alpha'}(\theta)
\]
is concave or convex over the appropriate region.  In the remainder
of this section, we give the results of computing (numerically) the values 
of this function $I_{\alpha'}$.
For small enough $\alpha$ it is convex,
so that we need two values of $\alpha'$, corresponding to
a POVM with six outcomes.

\begin{figure}[tbp]
\vspace*{1.4in}
\begin{center}
\epsfig{height=3in,file=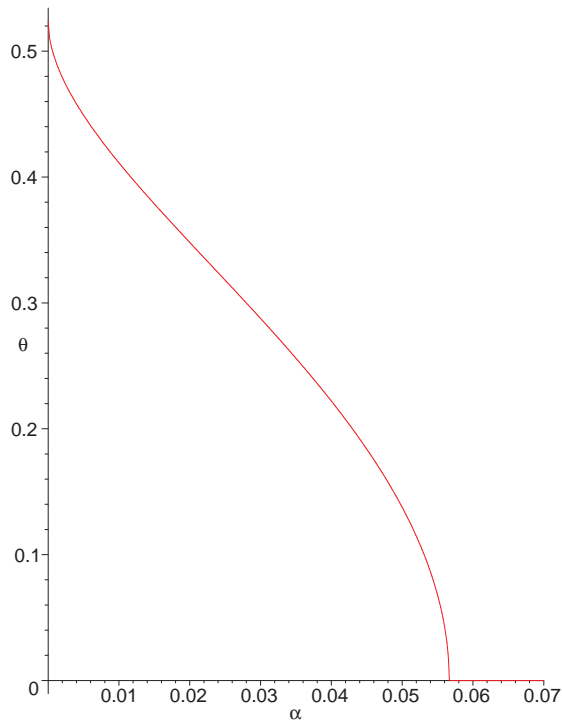}
\end{center}
\caption{The value of $\theta$ maximizing $I_\alpha$
for $\alpha$ between $0$ and $0.07$. This function starts at $\pi/6$ 
at $\alpha=0$, decreases until it hits $0$ at
$\alpha = 0.056651$ and stays at $0$ for larger $\alpha$.
\label{fig-opttheta}
}
\end{figure}

We need to calculate the Shannon capacity of the classical channel whose
input is one of the three trine states $T(\alpha')$, and whose output is 
determined by the von Neumann measurement $V(\theta)$.  Because of
the symmetry, 
we can calculate this using only the first projector $V_0$.
The Shannon mutual information between the input and the output is 
$H(X_\mathrm{in}) - H(X_\mathrm{in} | X_\mathrm{out})$,
which is
\begin{equation}
I_{\alpha'} = \log_2 3 +\sum_{b=0}^2 
\left|\braket{V_0(\theta)}{T_b(\alpha')}\right| ^2 
\log_2 \left|\braket{V_0(\theta)}{T_b(\alpha')}\right|^2 .
\end{equation}
The $\theta$ giving the maximum $I_\alpha'$ is $\pi/6$ for
$\alpha' = 0$,  
decreases continuously to 0 at $\alpha' = 0.056651$, and
remains 0 for larger $\alpha'$.  (See Fig. \ref{fig-opttheta}.)  
This value $0.056651$ corresponds to an angle
of 0.24032 radians ($13.769^\circ$).  This $\theta$ was determined
by using the computer package Maple 
to numerically find the point at which $d I_\alpha(\theta)/ d\theta = 0$.

\begin{figure}[tpb]
\vspace*{1.4in}
\begin{center}
\epsfig{file=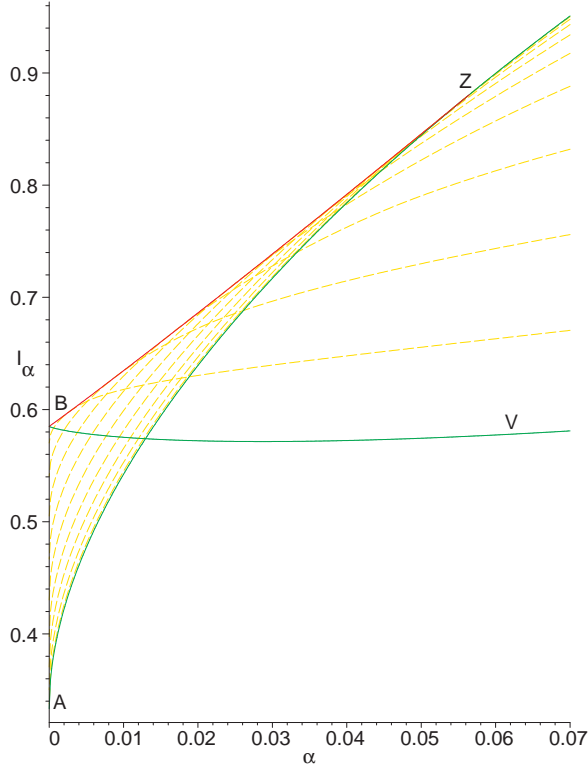,height=3in}
\end{center}
\caption{This plot shows $I_\alpha(\theta)$ for $0 \leq \alpha \leq 0.07$ 
and various $\theta$.  This is the mutual information between the
lifted trines at an angle of $\arcsin \sqrt{\alpha}$ to the $x$-$y$ plane,
and a von Neumann measurement rotated with respect to these trines
by an angle $\theta$.
The green curve AZ is $I_\alpha(0)$ 
and the green curve BV is $I_\alpha(\pi/6)$.  
The $\theta = 0$ curve is optimal
for $\alpha>0.056651$, and $\theta = \pi/6$ is
optimal for $\alpha=0$.  The dashed yellow curves show
$I_\alpha(\theta)$ for $\theta$ at intervals of $3^\circ$ between
$0^\circ$ and $30^\circ$ ($\pi/6$ radians).
Finally, the red curve BZ shows $I_\alpha(\theta_{\rm opt})$ for those $\alpha$
where neither $0$ nor $\pi/6$ is the optimal $\theta$.  The function
$\theta_{\mathrm{opt}}$ is given in Figure~\protect{\ref{fig-opttheta}}.
It is hard to
see from this plot, but the red curve BZ is slightly convex, i.e., its second
derivative is positive.  
This is clearer in Fig.~\protect{\ref{fig-accessible}}.
\label{fig-manythetas}
}
\end{figure}

By plugging the optimum $\theta$ into the formula for $I_{\alpha'}$, we 
obtain the optimum von Neumann measurement of the form $V$ above.  
We believe this is also the optimal generic von Neumann measurement,
but have not proved this.  
The maximum of $I_{\alpha'}(\theta)$ over $\theta$, and curves 
that show the behavior of $I_{\alpha'}(\theta)$
for constant $\theta$, are plotted in Fig.~\ref{fig-manythetas}.  
We can now observe that the leftmost piece of the curve is convex, and thus
that for small $\alpha$ 
the best POVM will have six projectors, corresponding to two values
of $\alpha'$.  For trine states
with $0 < \alpha < 0.061367$, the two values of $\alpha'$ giving the 
maximum accessible information are
$0$ and $0.061367$; we call this second value $\gamma_1$.
The trine states $T(\gamma_1)$ make an angle of
0.25033 radians ($14.343^\circ$) with the $x$-$y$ plane.

\begin{figure}[tbp]
\vspace*{1.4in}
\begin{center}
\epsfig{height=3in,file=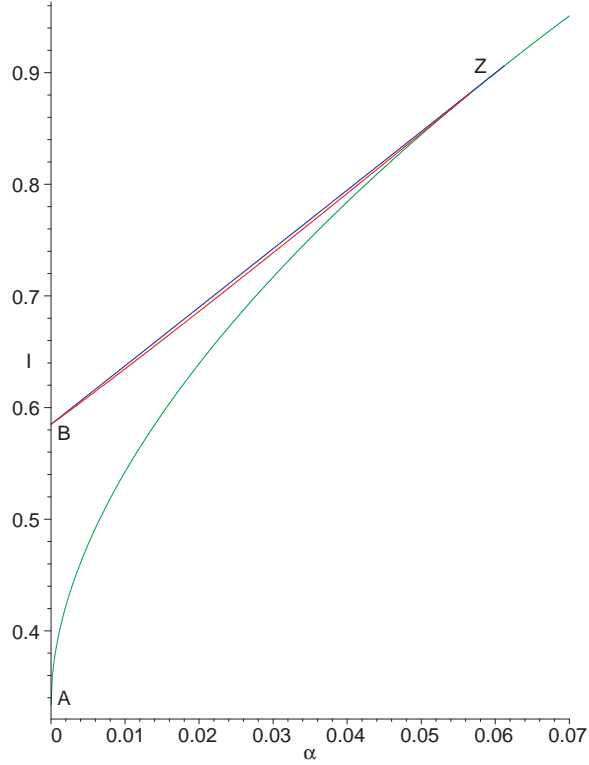}
\end{center}
\caption{This graph contains three curves.  
As in Fig.~{\protect{\ref{fig-manythetas}}},
the green curve AZ is $I_\alpha(0)$ and the red curve BZ is the maximum over
$\theta$ of $I_\alpha(\theta)$ for $\alpha < 0.056651$ (for larger
$\alpha$, this maximum is the green curve). 
The blue line BZ is straight; it is 
the convex envelope of the red and green curves and lies slightly above
the red curve BZ.  This blue line is a
linear interpolation between $\alpha = 0$ and $\alpha = 0.061367$ 
and corresponds to a POVM having six elements.
It gives the accessible information for the lifted trine states $T(\alpha)$
when $0 \leq \alpha \leq 0.061367$.  The difference between the blue
and red curves is maximum at $\alpha = 0.024831$, when this difference
reaches $0.0038282$. 
\label{fig-accessible}
}
\end{figure}

We can now invert the formula for $\alpha'$ (Eq.~\ref{alpha-prime}) 
to obtain a formula for
$\sin^2 \phi$, and substitute the value of 
$\alpha'= \gamma_1$ back into the formula 
to obtain the optimal POVM.  We find
\begin{eqnarray}
\nonumber
\sin^2(\phi_\alpha) &=& 
\frac{1-\alpha}{1+\alpha\left(\frac{2-3\gamma_1}{\gamma_1}\right)}\\
&\approx& \frac{1-\alpha}{1+29.591\alpha}
\label{phi-formula}
\end{eqnarray}
where $\gamma_1 \approx 0.061367$ as above.  
Thus, the elements in the optimal POVM we have found
for the trines $T(\alpha)$, 
when $\alpha < \gamma_1$,
are the six vectors
$P_b(\phi_\alpha, 0)$ and $P_b(0, \pi/6)$, where $\phi_\alpha$ is given by
Eq.~\ref{phi-formula} and $b= 0, 1, 2$.  
Fig.~\ref{fig-accessible} plots the accessible information given by
this six-outcome POVM, and compares it to the accessible information
obtained by the best known von Neumann measurement.

We also prove there are no other POVM's which attain the same
accessible information.  
The argument above shows that any optimal POVM must
contain only projectors chosen from these six vectors:
only those two values of $\alpha'$ can
appear in the measurement giving maximum capacity, and for each of 
these values of $\alpha'$ there 
are only three projectors in $V(\theta)$ which can
maximize $I_{\alpha'}$ for
these $\alpha'$.  It is easy to check that there is only one set of 
probabilities $p_i$ which make the above six vectors into a POVM, 
and that none of these probabilities are 0 for $0 < \alpha < \gamma_1$.  
Thus, for the lifted trine states with $0 < \alpha < \gamma_1$, 
there is only one POVM maximizing accessible information, and it 
contains six elements, the maximum possible for real states by a
generalization of Davies' theorem \cite{Davies-gen}.

\section{The $\mathbf{C_{1,1}}$ Capacity}
\label{C11-sec}
In this section, we discuss
the $C_{1,1}$ capacity (or one-shot capacity)
of the lifted trine states.  This is the maximum
of the accessible information over all probability distributions of
the lifted trine states.
Because the trine states are real vectors, it follows from a version of
Davies' theorem that there is an optimal POVM
with at most six elements, all the components of which are real.  
Since the lifted trine states are three-fold symmetric, one might
expect that the solution maximizing $C_{1,1}$ capacity is also
three-fold symmetric.  However, unlike
accessible information, for $C_{1,1}$ capacity
a symmetric problem does not mean that the optimal 
probabilities and the optimal measurement can be made symmetric.  Indeed, 
for the planar trine states, it is known that they
cannot.  The optimal
$C_{1,1}$ capacity for the planar trine states $T(0)$ is obtained by
assigning probability $1/2$ 
to two of the three states and not using the third one at all.  
(See Appendix A.)  This
gives a channel capacity of $1-H( 1/2-\sqrt{3}/4)= 0.64542$ bits, where 
$H(x) = -x \log_2 x - (1-x) \log_2 (1-x) $ 
is the binary Shannon entropy.  As discussed in the previous section,
the accessible information when all three trine states have equal 
probability is $\log_2 3 -1 = 0.58496$ bits.

In this section, we first discuss the best measurement we have found
to date.  We believe this is likely to be the optimal measurement, but
do not have a proof of this.  Later,
we will discuss what we can actually prove; namely,
that as $\alpha$ approaches 0 (i.e., for nearly planar trine states), the
actual $C_{1,1}$ capacity becomes exponentially close to the value given by
our conjectured optimal measurement.  We postpone this proof to 
Section~\ref{sec-exp-close} so that in Section~\ref{sec-adaptive}
we can complete our presentation of the various channel
capacities of the lifted 
trine states by giving an adaptive protocol that improves on our conjectured 
$C_{1,1}$ capacity.  Together with the bounds in Section~\ref{sec-exp-close},
this lets us prove that the adaptive capacity $C_{1,A}$ is strictly 
larger than $C_{1,1}$.  

Our starting point is the $C_{1,1}$ capacity for planar trines.  The
optimum probability distribution uses 
just two of the three trines.  For two pure states,
$\ket{v_1}$ and $\ket{v_2}$,
the optimum measurement for $C_{1,1}$ is known.  Let the states 
have an angle $\theta$ between them, so that
$\big|\braket{v_1}{v_2}\big|^2 = \cos^2 \theta$.  We can then take the two 
states to be
$v_1 = (\cos \frac{\theta}{2}, \sin \frac{\theta}{2})$ and
$v_2 = (\cos \frac{\theta}{2}, - \sin \frac{\theta}{2})$.  
The optimal measurement is the von Neumann measurement with projectors
$P{\pm} = (1/\sqrt{2}, \pm 1/\sqrt{2})$.
This measurement induces a classical binary symmetric channel with
error probability 
\begin{eqnarray*}
\braket{P_+}{v_2}^2 &=&\cos^2 (\theta/2+\pi/4)\\
&=& \frac{1-\sin \theta}{2}.
\end{eqnarray*}
and the $C_{1,1}$ capacity is thus $1-H(\frac{1}{2} -\frac{1}{2} \sin \theta)$.
Thus, for the planar trines, the $C_{1,1}$ capacity is 
$1-H( 1/2-\sqrt{3}/4)= 0.64542$.
To obtain our best guess for the $C_{1,1}$ capacity of the lifted trines with 
$\alpha$ small, 
we will give three
successively better guesses at the optimal probability distribution
and measurement.  
For small $\alpha$, we know of nothing better than the third
guess, which we conjecture to be optimal when $\alpha < 0.018073$.  
John Smolin has tried searching for solutions using a 
hill-climbing optimization program, and failed to find any better 
measurement for $C_{1,1}$, although the program did converge to the best 
known value a significant fraction of the time \cite{smolin-pc}.

\begin{figure}[htbp]
\vspace*{1.4in}
\begin{center}
\epsfig{height=3in,file=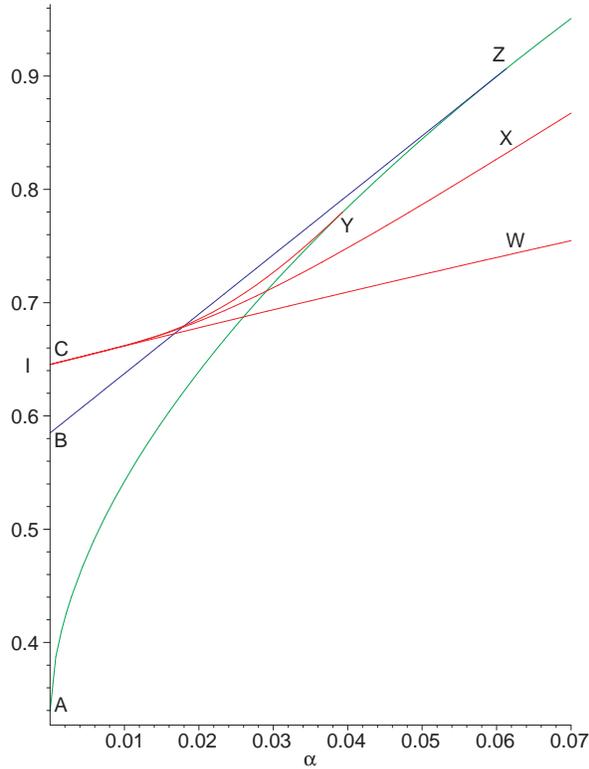}
\end{center}
\caption{\small This graph contains five curves.  
The blue line BZ and the green curve AYZ are the same as in 
Fig.~{\protect{\ref{fig-accessible}}}.  The maximum of these two 
curves is the accessible information for the lifted trine states 
$T(\alpha)$ with equal probabilities for all three states.  The maximum of
all five curves is the conjectured $C_{1,1}$ capacity.  
The three red curves with left endpoint C are the three
successively better guesses 
described in the text
for the $C_{1,1}$ capacity.
The lower red curve CW is the $C_{1,1}$ capacity for 
just two of the lifted trine states.  The middle red curve CX
is the capacity obtained using the same measurement 
$Q( 4\alpha / (3\alpha+1))$  
that gives the lower red curve CW, but with the probabilities of the three
trine states optimized.  Finally, the top red curve CY
is the $C_{1,1}$ capacity obtained by optimizing both the probabilities
and the measurement, but only over the limited class of measurements $Q(b)$.  
These three red curves become exponentially close to each other 
as $\alpha$ approaches~0.  
\label{fig-c1}
}
\end{figure}

For the trines $T(0)$, the optimum probability distribution is 
$(\frac{1}{2},\frac{1}{2},0)$.  
Our first guess is to continue to use the same probability distribution 
for $\alpha > 0$.  For the trines $T(\alpha)$, this probability 
distribution, $(\frac{1}{2},\frac{1}{2},0)$,  
the optimum measurement is a
von Neumann measurement with projectors
\begin{eqnarray}
\nonumber
Q_0(\beta) &=& \ \left(\sqrt{\beta}, 0, \sqrt{1-\beta}\,\right) \\ \nonumber
Q_1(\beta) &=& \frac{1}{\sqrt{2}} \left(-\sqrt{1-\beta}, 1, \sqrt{\beta}\,\right) \\
Q_2(\beta) &=& \frac{1}{\sqrt{2}} \left(-\sqrt{1-\beta}, -1, \sqrt{\beta}\,\right)
\label{def-Q}
\end{eqnarray}
where $\beta = 4\alpha / (3\alpha+1)$.  
The $C_{1,1}$ capacity in this case is $1-H(p)$ where 
$p = \frac{1}{2}(3\alpha-1)^2/(3\alpha+1)$.  This function is plotted  
in Fig.~\ref{fig-c1}.  We call this the two-trine capacity.

\begin{figure}[tbp]
\begin{center}
\phantom{,}\epsfig{height=2in,file=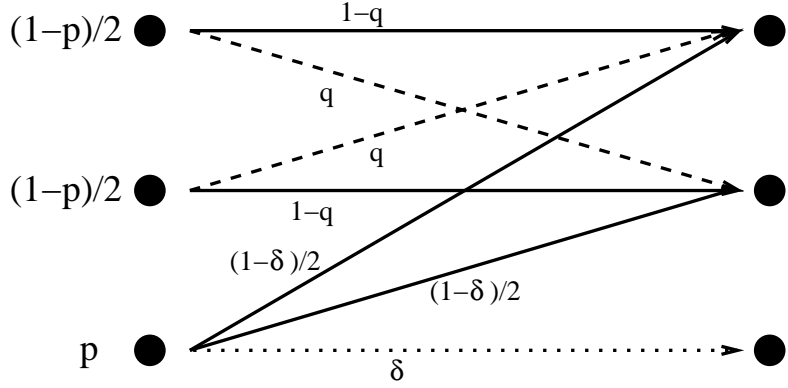}
\end{center}
\caption{\small 
The classical channel induced by the measurement $Q(4\alpha/(3\alpha+1))$
on the lifted trines $T(\alpha)$.  The inputs, from top to bottom,
correspond to $T_1$, $T_2$, and $T_0$;  the outputs correspond to
$Q_1$, $Q_2$ and $Q_0$.  The transition probabilities are given above,
where $\delta = \braket{Q_0}{T_0}^2$ and $q = \braket{Q_1}{T_2}^2 = 
\braket{Q_2}{T_1}^2$.
\label{fig-channel-epsilon-0}
}
\end{figure}
The second guess comes from using the same measurement, $Q(\beta)$,
as the first guess, 
but varying the probabilities of the three trine states
so as to maximize the $C_{1,1}$ capacity obtained
using this measurement.  To do this, we need to consider the classical
channel shown in Fig.~\ref{fig-channel-epsilon-0}.  Because of the symmetry
of this channel, the optimal probability distribution is guaranteed to give
equal probabilities to trines $T_1$ and $T_2$.  Remarkably, this
channel has a closed form for the probability $p$ for the third trine
which maximizes the mutual information.  Expressing the mutual information
as $H(X_{\mathrm{out}}) - H(X_{\mathrm{out}}|X_{\mathrm{in}})$,
we find that this simplifies to
\begin{equation}
(1-p)(1-H(q)) + H(p\delta) -p H(\delta) 
\end{equation}
where $\delta = \braket{Q_0}{T_0}^2$ and 
$q = \braket{Q_2}{T_1}^2 = \braket{Q_1}{T_2}^2$.  
Taking the derivative of this function with respect to $p$, 
setting it to 0, and moving the terms with $p$ in them to the left
side of the equality gives
\begin{equation}
\delta \left[ \log_2(1-p\delta) - \log_2(p)\right] = 
1-H(q) -  (1-\delta)\log_2(1-\delta).
\end{equation}
Dividing by $\delta$ and exponentiating both sides gives
\begin{equation}
\frac{1-\delta p}{p} = 
2^{\textstyle \, \frac{1}{\delta} \left({1-H(q) - (1-\delta)\log_2(1-\delta)} \right)}
\end{equation}
which has the solution
\begin{equation}
p = \frac{1}{\delta  + \exp \left( \frac{\log 2}{\delta}  [ 1-H(q) - 
(1-\delta)\log_2(1-\delta)] \right)} .
\end{equation}
Using this value of $p$, and the measurement of Eq.~\ref{def-Q} with
$\beta = 4\alpha / (3\alpha+1)$, we obtain a curve that is plotted in
Fig.~\ref{fig-c1}.  Note that as $\alpha$ goes to 0, $\delta$ goes to~0 
and the exponential
on the right side goes to $2^{[1-H(q)]/\delta}$, so
$p$ becomes exponentially small.  It follows that this function differs
from the two-trine capacity by an exponentially small amount as $\alpha$
approaches $0$.  Note also that no matter how small $\delta$ is,
the above value of $p$ is non-zero, so even though the two-trine capacity
is exponentially close to the above capacity, it is not equal.

\begin{figure}[tbp]
\epsfysize=2in
\begin{center}
\phantom{1}\epsfig{height=2in,file=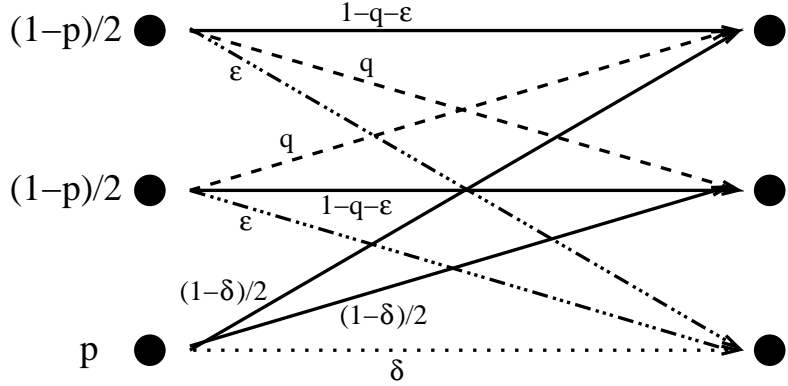}
\end{center}
\caption{\small 
The classical channel induced by the von Neumann
measurement $Q(\beta)$ on the lifted trines $T(\alpha)$. 
The inputs correspond (top to bottom) to $T_1$, $T_2$ and $T_0$; the 
outputs correspond (top to bottom) to $Q_1$, $Q_2$ and $Q_0$.
\label{fig-channel-epsilon-positive}
}
\end{figure}

\begin{figure}[tbp]
\vspace*{1.5in}
\begin{center}
\phantom{,}\epsfig{height=3in,file=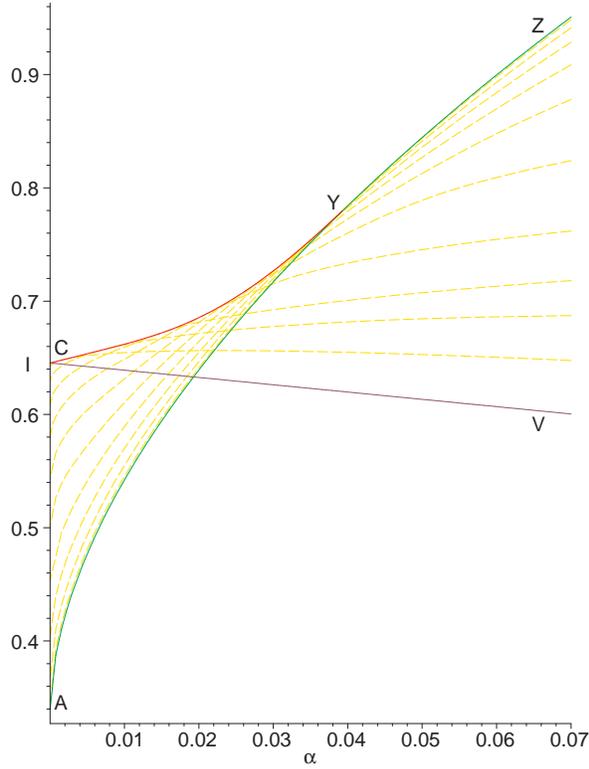}
\end{center}
\caption{\small
This figure shows the Shannon capacity obtained using 
various measurements on the trine states $T(\alpha)$, while optimizing
the input probability distribution on the trines.  The green curve AYZ
is the same as in the previous figures.  It is obtained using the von 
Neumann measurement $Q(2/3)$, which is also
the measurement $V(0)$. The violet curve CV is 
obtained using the measurement $Q(0)$, which is optimal
for the planar trines ($\alpha = 0$).  The dashed yellow curves are the 
capacities obtained by the measurement $Q(\sin^2 \theta)$ where $\theta$ 
is taken at intervals of $5^\circ$ from $5^\circ$ 
to $50^\circ$.  The violet curve CV corresponds to $\theta = 0^\circ$ 
and the green curve AYZ to $\theta = 54.736^\circ$.
Finally, the red curve CY is the upper envelope of the dashed yellow
curves; shows the capacity obtained by choosing the 
measurement $Q(\beta)$ with optimal $\beta$ for each $\alpha < 0.040491$.  
For larger $\alpha$, this optimum is at $\beta=2/3$, and is given by the
green curve YZ.
\label{fig-q1}
}
\end{figure}

For our third guess, we refine the above solution slightly.  It turns out
that the $\beta$ used to determine the measurement $Q(\beta)$ is no longer 
optimal after we have given a non-zero probability to the third
trine state.  What we do is vary both $p$ and $\beta$ to find the optimal
measurement for a given $\alpha$.  This leads to the classical channel
shown in Fig.~\ref{fig-channel-epsilon-positive}.  Here, 
$q$ and $\delta$ take the same values as above, and 
$\epsilon = \braket{Q_0}{T_1}^2 = \braket{Q_0}{T_2}^2$.
As we did for the case with $\epsilon = 0$, we can write down the channel
capacity, differentiate with respect to $p$, and solve the resulting 
equation.  In this case, the solution turns out to be
\begin{equation}
p = \frac{ 1 - \epsilon - \epsilon 2^Z}{ (\delta - \epsilon)(1+2^Z)}
\label{opt-p}
\end{equation}
where
\begin{equation}
Z = \frac{ 1 - \epsilon - H(q;\epsilon;1-q-\epsilon) 
+H(\delta)}{\delta-\epsilon}.
\label{opt-Z}
\end{equation}
Here 
\[
H(p_1; p_2; \cdots; p_k) = \sum_{j=1}^k -p_j \log_2 p_j
\]
is the Shannon entropy of the 
probability distribution $\{p_1, p_2, \cdots, p_k\}$.
We have numerically found the optimum $\beta$ for the measurement $Q(\beta)$,
and used this result to obtain the $C_{1,1}$ capacity
achieved by optimizing over both $p$ and $Q(\beta)$.  This capacity
function is shown in Fig.~\ref{fig-c1}.  This capacity, and the capacities 
obtained using various specific values of $\beta$ in $Q(\beta)$ are shown 
in Fig.~\ref{fig-q1}.
For $\alpha \geq 0.040491$, the optimum $\beta$ is $\frac{2}{3}$;
note that the measurement $Q(\frac{2}{3})$ is the same
as the measurement $V(0)$ introduced in Section~\ref{sec-accessible}, 
Eq.~\ref{defineV}.
The measurement $Q(\beta)$ appears to give the $C_{1,1}$ capacity for
$\alpha \leq 0.018073$ [and for $\alpha \geq 0.061367$, where the optimum
measurement is $Q(\frac{2}{3})$].

Now, suppose that in the above expression for $p$ [Eqs.~(\ref{opt-p}) and
(\ref{opt-Z})], 
$\epsilon$ and $\delta$ are 
both approaching $0$, while $q$ is bounded away from $\frac{1}{2}$.
If $\delta > \epsilon$, then $2^Z$ is exponentially large
in $1/(\delta - \epsilon)$, and the equation 
either gives a negative $p$  (in which case the optimum value of $p$ is 
actually $0$)
or $p$ is exponentially small.
If $\epsilon > \delta$, and both 
$\epsilon$ and $\delta$ are sufficiently small, then $2^Z$ is exponentially 
small in $1/(\epsilon - \delta)$ and the value of $p$ in the above 
equation is negative, so that
the optimum value of $p$ is $0$.  There are solutions to the above equation
which have $\epsilon > \delta$ and positive $p$, but this is not the case
when $\epsilon$ is sufficiently close to $0$.

It follows from the above argument that, as $\alpha$ goes to $0$, the
optimum probability $p$ for the third trine state
goes to $0$ exponentially fast in $1/\delta$, and so the $C_{1,1}$ 
capacity obtained
from this measurement grows exponentially close to that for the two-trine
capacity, since the two probability distributions differ by an exponentially
small amount.   

We have now described our conjectured $C_{1,1}$ for the lifted trine states,
and the measurements and probability distributions that achieve it.
In the next section, we will show that there is an adaptive protocol
which achieves a 
capacity $C_{1,A}$ considerably better than our conjectured $C_{1,1}$
capacity.  
When $\alpha$ is close to $0$, it is better by an amount linear 
in $\alpha$. To rigorously prove that it is better, we need to find an 
upper bound on the capacity $C_{1,1}$ which is less than $C_{1,A}$.  
We already noted that, as
$\alpha$ approaches $0$, all three of our guesses for $C_{1,1}$ become 
exponentially close.  
In Section~\ref{sec-exp-close} 
of this paper we prove that the 
true $C_{1,1}$ capacity must
become exponentially close to these guesses.  

Because these three guesses for $C_{1,1}$ become exponentially close
near $\alpha=0$, they all have the same derivative with respect to
$\alpha$ at $\alpha=0$.  Our first guess, which used only two of the three
trines, is simple enough that we can
compute this derivative analytically, and we find that its value is
\[
\frac{\sqrt{3}}{2} \log_2\left( 2 + \sqrt{3} \right) = 1.64542 
{\hbox{\ bits.}}
\]
This contrasts with our best adaptive protocol, which has 
the same capacity 
at $\alpha=0$, but which between $\alpha=0$ and $\alpha=0.087247$
has slope 4.42238 bits (see Fig.~\ref{fig-adapt-narrow}).  
Thus, for small enough $\alpha$, the adaptive capacity $C_{1,A}$ is strictly 
larger than $C_{1,1}$.

\newpage

\section{The Adaptive Capacity $\mathbf C_{1,A}$}
\label{sec-adaptive}

As can be seen from Figure~\ref{fig-c1}, 
the $C_{1,1}$ capacity is not concave in $\alpha$.  
That is, there are two values of $\alpha$ such that the
average of their $C_{1,1}$ capacities is larger than the $C_{1,1}$ capacity
of their average.  This is analogous to the situation we found while studying
the accessible information for the probability distribution 
$( \frac{1}{3},\frac{1}{3},\frac{1}{3})$, where the curve giving the
information attainable by von 
Neumann measurements was also not concave.  In that case, we were able to 
obtain the convex hull of this curve by using a POVM to linearly
interpolate between the two von Neumann measurements.  Remarkably,
we show that for the lifted trines example, 
the relationship between $C_{1,1}$ capacity and $C_{1,A}$ capacity
is similar: 
protocols using adaptive measurement can attain the convex hull of
the $C_{1,1}$ capacity with respect to $\alpha$.  

We now introduce the adaptive measurement model leading to the
$C_{1,A}$ capacity.  
If we assume that each of the signals that Bob receives is held by a 
separate party, this is the same as the LOCC model used in
\cite{Horodecki-be,unmeasurable-product} where several parties share
a quantum state and are allowed to use local quantum operations
and classical communication between the parties.
In our model, Alice sends
Bob a tensor product codeword using the channel many times.
We call the output from a single use of the channel a {\em signal}.
Bob is not allowed to
make joint quantum measurements on more than one signal, but
he is allowed to make measurements sequentially on different 
signals.  
He is further
allowed to use the classical outcomes of his measurements 
to determine which signal to measure next, and to determine which 
measurement to make on that signal.  
In particular, he is allowed to make a 
measurement which only partially reduces the quantum state of one signal, 
make intervening measurements on other signals, and return to make
a further measurement on the reduced state of the original signal (which
measurement may depend on the outcomes of intervening measurements).
The information rate for a given 
encoding and measurement strategy
is the mutual information between
Alice's codewords and Bob's measurement outcomes, divided by the number
of signals (channel uses) in the codeword.  The adaptive one-shot capacity 
$C_{1,A}$ is defined to be the supremum of this information rate
over all encodings and all measurement
strategies that use quantum operations local to the separate
signals (and classical computation to coordinate them).  
As we will show in Section~\ref{sec-discussion}, to exceed $C_{1,1}$ it
is crucial to be able to refine a measurement made on a given signal
after making intervening measurements on other signals.

In our adaptive protocol for lifted trines, we use two 
rounds of measurements.  
We first make one measurement on each of the signals received; 
this measurement only partially reduces the quantum state of some 
of the signals.  We then make a second 
measurement (on some of the signals) which depends on the 
outcomes of the first round of measurements.

A precursor to this type of adaptive strategy appeared in an 
influential paper of Peres and Wootters \cite{Peres-Wootters}
which was a source of inspiration for this paper.
In their paper, Peres and Wootters
studied strategies of adaptive measurement on the tensor product 
of two trine states, in which joint measurements on both
copies were not allowed.  They showed that for a 
specific encoding of block length 2, adaptive strategies could extract
strictly more information than
sequential strategies, but not as much as was
extractable through joint measurements.  
However, the adaptive strategies they considered extracted less
information than the $C_{1,1}$ capacity of the trine states, and so 
could have been improved on by using a different encoding and a sequential
strategy.  We show that for some values of $\alpha$, 
$C_{1,A}$ is strictly greater than $C_{1,1}$ for the 
lifted trines $T(\alpha)$, where these capacities are defined using 
arbitrarily large block lengths and arbitrary encodings.  
It is open whether $C_{1,1} = C_{1,A}$ for the planar trine states.

Before we describe our measurement strategy, we will describe the codewords
we use for information transmission.  The reason we choose these
codewords will not become clear until after we have given the strategy.
Alice will send one of these codewords
to Bob, who with high probability will be able to deduce which codeword
was sent from the outcomes of his measurements.  These
codewords are constructed
using a two-stage scheme, corresponding to the two rounds of our measurement
protocol.  Effectively, we are
applying Shannon's classical channel coding theorem twice.  

To construct a codeword, we take two error correcting 
codes each of block length $n$ and 
add them letterwise (mod 3). The first code is over a trinary alphabet 
(which we take 
to be $\{0,1,2\})$; it contains $2^{\tau_1 n-o(n)}$ codewords, and is a
good classical error-correcting code for a classical channel we describe later.  
The second code contains $2^{\tau_2n-o(n)}$ codewords,
is over the binary alphabet $\{0,1\}$, and is a good classical
error-correcting code for a different classical channel.   
Such classical error-correcting codes
can be constructed by taking the appropriate number of random 
codewords; for the proof that
our decoding strategy works,
we will assume that the codes were indeed
constructed this way. 
To obtain our code, we simply add these two codes bit-wise (mod 3).  
For example, if a codeword in the first (trinary) code is 0212 and a 
codeword in the second (binary) code is 1110, the codeword obtained 
by adding them bitwise is 1022. 
This new code contains $2^{(\tau_1 + \tau_2)n - o(n)}$ codewords
(since we choose the two codes randomly, and we make sure
that $\tau_1 + \tau_2 < \log_2 3$).
 
To show how our construction works, 
we first consider the following measurement strategy.  This is 
not the best protocol we have found, but it provides a good illustration
of how our protocols work.
This uses the two-level codeword scheme described above.
In this protocol, the first measurement we make uses a POVM which 
contains four elements.  
One of them is a scalar times the matrix 
\[
\Pi_{xy} = \left(
\begin{array}{ccc}
1 & 0 & 0 \\
0 & 1 & 0 \\
0 & 0 & 0 
\end{array}
\right),
\]
which projects the lifted trine states onto the planar trine states.
The other three elements correspond to the three
vectors which are each perpendicular to two of the trine states; these
vectors are
\begin{eqnarray*}
D_0 &= & \left(\frac{2\sqrt {\alpha}}{\sqrt{1+3\alpha}}, 0, 
\frac{\sqrt {1-\alpha}}{\sqrt{1+3\alpha}}\right)\\
D_1 & = & \left(-\frac{\sqrt {\alpha}}{\sqrt{1+3\alpha}},  
  \frac{\sqrt {3\alpha}}{\sqrt{1+3\alpha}},  
 \frac{\sqrt {1-\alpha}}{\sqrt{1+3\alpha}}\right)\\
D_2 & = & \left(-\frac{\sqrt {\alpha}}{\sqrt{1+3\alpha}},  
  -\frac{\sqrt {3\alpha}}{\sqrt{1+3\alpha}},  
 \frac{\sqrt {1-\alpha}}{\sqrt{1+3\alpha}}\right).
\end{eqnarray*}
Note that $\braket{D_{b_1}}{T_{b_2}} = 0$ if and only if $b_1 \neq b_2$.

We now scale $\Pi_{xy}$ and $D_i$ so as to make a valid POVM.
For this, we the POVM elements to sum to the identity, i.e., that
$\sum_{i=0}^3 A_i^\dag A_i = I$.  This is 
done by choosing
\begin{eqnarray*}
A_i &=& \frac{\sqrt{1+3\alpha}}{\sqrt{3(1-\alpha)}}\;\proj{D_i}, 
\qquad \qquad i=0,1,2 \\
A_3 &=& \frac{\sqrt{1-3\alpha}}{\sqrt{1-\alpha}}\:\Pi_{xy}
\end{eqnarray*}
When we apply this POVM, 
the state $\ket{v}$ is taken to the state $A_i\ket{v}$ with probability
$\bra{v} A_i^\dag A_i \ket{v}$.
When this operator is applied to a trine state $T_b(\alpha)$, the chance of 
obtaining $D_b$ is $3 \alpha$, and the chance of obtaining $\Pi_{xy}$ 
is $1-3\alpha$.  
If the outcome is $D_b$, we know we started with the trine $T_b$,
since the other two trines are perpendicular to $D_b$.  If we obtain
the fourth outcome, $\Pi_{xy}$, then we gain no information about which of the 
three trine states we started with, since all three states are equally
likely to produce $\Pi_{xy}$.  

\begin{figure}[tbp]
\epsfysize=2in
\begin{center}
\phantom{1}
\phantom{,}\epsfig{height=2in,file=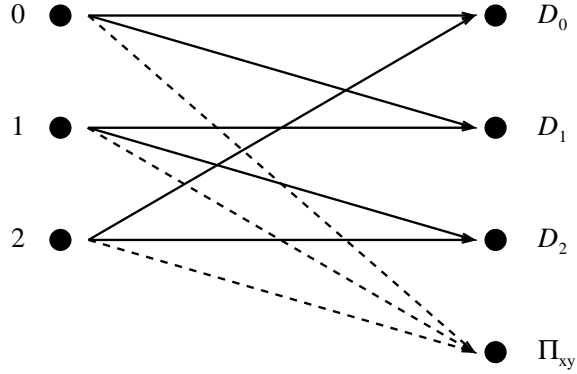}
\end{center}
\caption{\small 
The classical channel corresponding to the first-stage code in our first
adaptive protocol.
The solid lines indicate a probability of $3\alpha/2$ for the transition; 
dashed lines a probability of $1-3\alpha$.  For example, a symbol
$0$ in the first
stage code is first encoded with probability $\frac{1}{2}$ each by trines
$T_0$ and $T_1$.  Considering the effects of the first  measurement, 
a symbol
$0$ in the first stage code is equally likely (probability $3\alpha/2$) 
to be taken to measurement outcomes $D_0$ and $D_1$, and is taken to 
outcome $\Pi_{xy}$ (which for this channel is essentially an erasure) 
with probability $1-3\alpha$.  
\label{channel-firststagecode}
}
\end{figure}

We now consider how this measurement
combines with our two-stage coding scheme introduced above.  
We first show that with high probability we can decode our first 
code correctly.  We then show that if we 
apply a further measurement to each of
the signals which had the outcome $\Pi_{xy}$, 
with high probability we can decode our second code correctly.

In our first measurement,
for each outcome of the type $D_b$ obtained, we know that the trine sent
was $T_b$.  However, this does not uniquely identify the letter $a$ in the
corresponding position of the 
first-stage code, as the trine $T_b$ sent was obtained by adding either 
$0$ or $1$ to $a$ (mod 3) to obtain $b$.   Thus, if we obtained the outcome
$D_b$, the corresponding symbol of our first-stage codeword is either
$b$ or $b-1$ (mod 3), and because the second code is a random code,
these two cases occur with equal probability.
This is illustrated in Figure~\ref{channel-firststagecode}; if the codeword
for the first stage code is $a$, then the outcome of the first
measurement will be $D_a$
with probability $3\alpha /2$, $D_{a+1\ ({\mathrm{mod}}\ 3)}$ 
with probability $3\alpha/2$, 
and $\Pi_{xy}$ with probability $1-3\alpha$.  This is a classical channel
with capacity $3\alpha (\log_2 3 - 1)$: with probability $3\alpha$, we obtain
an outcome $D_x$ for some $x$, and in this case  we get 
$(\log_2 3 - 1)$ bits of information about the first-stage codeword;
with probability $1-3\alpha$, we obtain $\Pi_{xy}$, which gives us no 
information about this codeword.  
By Shannon's classical channel coding 
theorem, we can now take $\tau_1 = 3 \alpha(\log_2 3 - 1)$, and choose a 
first-stage code with $2^{\tau_1 n - o(n)}$ codewords that is
an error-correcting code for the classical channel shown 
in Fig.~\ref{channel-firststagecode}.
Note that in this calculation, we 
are using the fact that the second-stage code is a random code, to say that
measurement outcomes
$D_a$ and $D_{a+1\ ({\mathrm{mod}}\ 3)}$ are equally likely.  

\begin{figure}[tbp]
\epsfysize=2in
\begin{center}
\phantom{1}
\phantom{,}\epsfig{height=2in,file=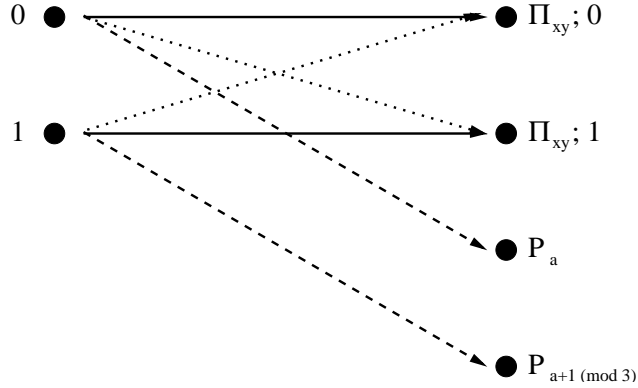}
\end{center}
\caption{\small 
The classical channel corresponding to the second-stage code in our first
adaptive protocol.
The solid lines indicate a probability of $0.64542 (1-3\alpha)$.  The
dotted lines indicate a probability of $0.35458 (1-3\alpha)$.  The
dashed lines indicate a probability of $3\alpha$.  Note that this channel
is symmetric if the inputs are interchanged; this means that the
probability distribution $(\frac{1}{2},\frac{1}{2})$ maximizes
the information transmission rate.
\label{channel-secondstagecode}
}
\end{figure}

Once we have decoded the first stage code, the uncertainty about which
codeword we sent is reduced to decoding the second stage code.
Because the second code is binary,
the decoding of the first-stage code leaves in each position
only two possible trines consistent with this decoding.
This means that in the approximately 
$(1 - 3\alpha) n$ positions where the trines are projected into the plane, 
we now need only distinguish between two of the three possible trines.  
In these positions, we can use the optimal measurement for distinguishing 
between two planar trine states; recall that this is a von Neumann measurement
which gives $1-H(1/2 + \sqrt{3}/4) = 0.64542$ bits of information per position.
In the approximately $3\alpha n$ remaining positions, we still know which
outcome $D_b$ was obtained in the first round of measurements, and
this outcome tells us exactly which trine was sent.  Decoding the 
first-stage code left us with two equally likely
possibilities in the second-stage code for each of these positions. 
We thus obtain one bit of information about the second-stage
code for each of these approximately $3\alpha$ positions.
Thus, if we set 
$\tau_2 = 0.64542(1-3\alpha) + 3 \alpha$ bits, by Shannon's theorem there is a
classical error-correcting code which can be used for the second stage
and which can almost certainly be decoded uniquely.  
Adding $\tau_1$ and $\tau_2$, we obtain a channel capacity of 
$0.64542(1-3 \alpha) + (\log_2 3) ( 3 \alpha)$ bits; this is the line
which interpolates between the points $\alpha=0$ and $\alpha = 1/3$
on the curve for $C_{1,1}$.  As can be seen from Figure \ref{fig-adapt-wide},
for small $\alpha$ this strategy indeed does 
better than our best protocol for $C_{1,1}$.

The above strategy can be viewed in a slightly different way, this time
as a three-step process.  In the first step our measurement either lifts 
the three trine states up until they are all orthogonal, or projects them 
into the plane.  This first step lifts approximately $3\alpha n$ trines up 
and projects approximately $(1-3\alpha) n$ trines into the plane.   After 
this first step, we first measure the trines that are lifted further out of
the plane, yielding
$\log_2 3$ bits of information for each of these approximately $3\alpha n$ 
positions.  The rest of the strategy then proceeds exactly as above.  Note
that this reinterpretation of the strategy is reminiscent of the 
two-stage description
of the six-outcome POVM for accessible information in 
Section~\ref{sec-accessible}.

\begin{figure}[tbp]
\vspace*{1in}
\begin{center}
\phantom{,}\epsfig{height=3in,file=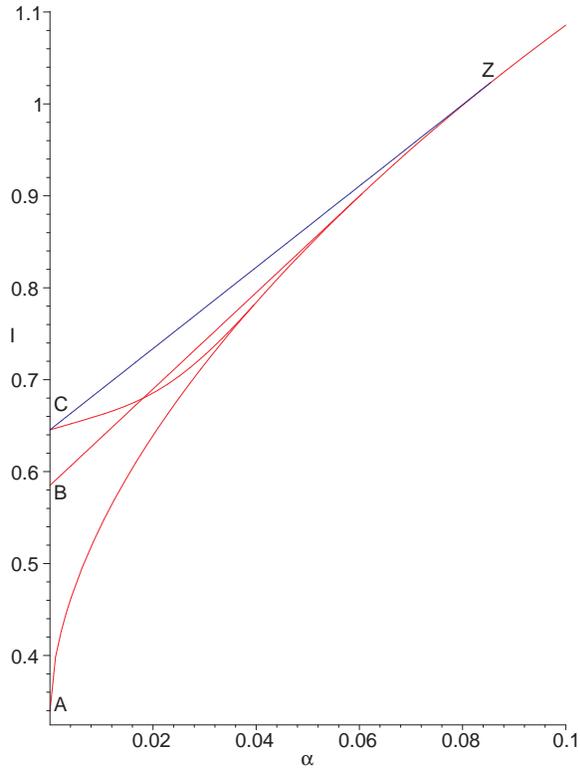}
\end{center}
\caption{\small 
The three red curves AZ, BZ and CZ 
are all shown in Fig.~\protect{\ref{fig-c1}}; their
maximum is the best value known for $C_{1,1}$  capacity.
The blue line CZ is straight; it is the second adaptive strategy 
discussed in this section, and is the largest value we know how 
to obtain for $C_{1,A}$.   
\label{fig-adapt-narrow}
}
\end{figure}

\begin{figure}[bt]
\vspace*{1.5in}
\begin{center}
\phantom{,}\epsfig{height=3in,file=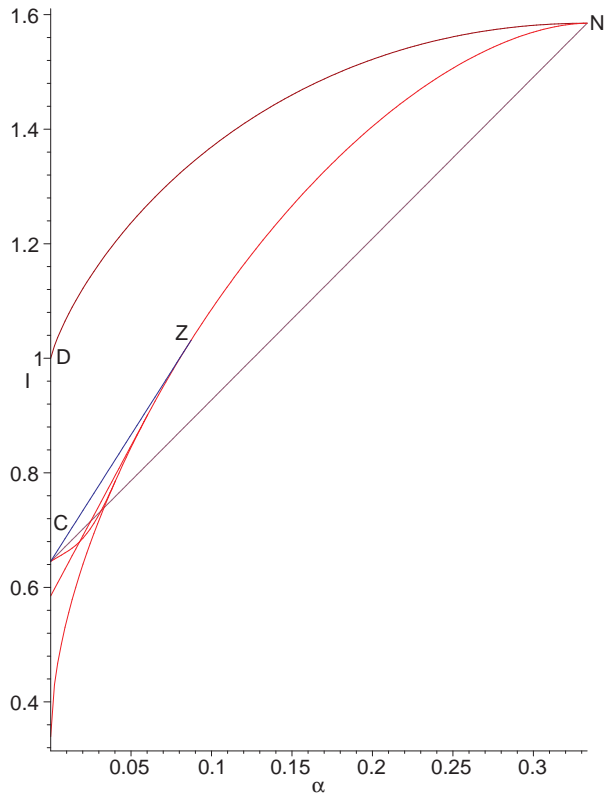}
\end{center}
\caption{\small 
All the curves in Figure \protect{\ref{fig-adapt-narrow}} are shown, for 
$0 \leq \alpha \leq \frac{1}{3}$, along with the purple line CN and the brown
curve DN.  The maximum of the three red curves is the best value for the
$C_{1,1}$ capacity
we have found.  The line CN is the capacity of the first
adaptive strategy discussed in in this section.  The line CZ is the 
capacity of the second adaptive strategy.  The curve DN is the 
$C_{1,\infty}$ capacity (i.e., the Holevo bound).
\label{fig-adapt-wide}
}
\end{figure}

We now modify the above protocol to give the best protocol we currently
know for the adaptive capacity $C_{1,A}$.  We first make a measurement 
which either projects the trines $T(\alpha)$ 
to the planar trines $T(0)$ or lifts
them out of the plane to some fixed height, yielding the trines
$T(\gamma_2)$; this measurement requires
$\alpha < \gamma_2$.  (Our first strategy is
obtained by setting $\gamma_2 = \frac{1}{3}$.)
We choose $\gamma_2 = 0.087247$; this is
the point where the convex hull of the curve
representing the $C_{1,1}$ capacity meets this curve 
(see Figure \ref{fig-adapt-narrow}); at this point $C_{1,1}$ 
is 1.03126 bits.  It is easy to verify that 
with probability $1-\alpha/\gamma_2$, the lifted trine
$T_b(\alpha)$ is projected onto a planar trine $T_b(0)$, and with 
probability $\alpha/\gamma_2$, it is lifted up to $T_b(\gamma_2)$.  
We next use the optimum von Neumann 
measurement $V(0) = Q(2/3)$ on the trine states that were lifted out of the 
plane.  

We now analyze this protocol in more detail.
Let 
\begin{equation}
p = \braket{V_b(0)}{T_b(\gamma_2)}^2 = 0.90364.
\end{equation}
The first-stage code gives an information gain of 
\[ \log_2 3 - H(\frac{1+p}{4}; \frac{1+p}{4}; \frac{1-p}{2}) = 0.35453 \ 
\mathrm{bits}
\]
for each of the approximately $(\alpha/\gamma_2) n$ 
signals which were lifted out of the plane.  This is because the
symbol $a$ in the first-stage code is first taken 
to $T_a$ or $T_{a+1\ ({\mathrm{mod}}\ 3)}$
with a probability of $\frac{1}{2}$ each (depending on the value of the
corresponding letter in the second-stage code).  We thus obtain 
each of the two
outcomes $V_a$, $V_{a+1\ ({\mathrm{mod}}\ 3)}$ with probabilities 
$\frac{1}{2}p + \frac{1}{2}(1-p)/2 = \frac{1}{4}(1+p)$, and obtain the
outcome $V_{a+2\ {(\mathrm{mod}}\ 3)}$ with probability 
$\frac{1}{2}(1-p)$.
Thus, if we start with the symbol $a$ in the first-stage code,
the entropy of the outcome of the measurement (i.e., 
$H(X_\mathrm{out} | X_\mathrm{in})$) is 
$H(\frac{1+p}{4}; \frac{1+p}{4}; \frac{1-p}{2})$; it is easy to see
that $H(X_\mathrm{out})$ is $\log_2 3$. 
From classical Shannon theory,
we find that we can take $\tau_1 = 0.35453(\alpha/\gamma_2)$.


The design of our codewords ensures that
knowledge of the first codeword eliminates 
one of the three signal states for each of the trines 
projected into the plane.  This allows us to use the optimal $C_{1,1}$ 
measurement on the planar trines, resulting in an information gain of
0.64542 bits for each of these approximately
$(1-\alpha/\gamma_2)n$ signals.  
We obtain 
\[
\frac{1}{2}(1+p)\left[1-H\left(\frac{1-p}{1+p}\right)\right] = 
0.67673 \quad \mathrm{bits} 
\]
for each of the approximately $\alpha/\gamma_2$
signals that were lifted out of the plane; this will be explained in more
detail later.  
Put together, this results in a capacity of 
\begin{equation}
C_{1,A} = 
0.64542 (1-\alpha/\gamma_2) + 1.03126(\alpha/\gamma_2)
\end{equation}
bits per signal; this
formula linearly interpolates between the $C_{1,1}$ capacity
for $T(0)$ and the $C_{1,1}$ capacity
for $T(\gamma_2)$.

Why do we get the weighted
average of $C_{1,1}$ for $T(0)$ and $C_{1,1}$ for $T(\gamma_2)$ as
the $C_{1,1}$ for $T(\alpha)$, $0 < \alpha < \gamma_2$?
This happens because we use all the information that
was extracted by both of the measurements.  The measurements on the trines
that were
projected onto the plane only give information about
the second code, and provide
0.64542 bits of information per trine.  For the trines $T(\gamma_2)$
that were lifted out of the plane, we use part of the information extracted
by their measurements to decode the first code, and part
to decode the second code.  For these trines,
in the first step, we start with the symbols $\{0,1,2\}$
of the first stage code with equal probabilities.  A 0 symbol 
gives measurement outcome $V_0$ with probability
$\frac{1+p}{4}$, $V_1$ with probability $\frac{1+p}{4}$, and 
$V_2$ with probability $\frac{1-p}{2}$, and similarly for the other signals.  
The information gain from this step is thus
\begin{equation}
\log_2 3 - H\left( {\textstyle 
\frac{1+p}{4};\frac{1+p}{4}; \frac{1-p}{2}} \right) = 0.35453 \quad 
\mathrm{bits} 
\label{info-stage1}
\end{equation}
per signal.  
At the start of the second step, we have 
narrowed the possible states for each signal
down to two equally likely possibilities.
This information gain for this step comes from the case where 
the outcome of the measurement was $V_b$, and where one of the two possible
states (consistent with the first-stage code)
is $T_b$.   This happens with probability $\frac{1}{2}(1+p)$.  
In this case, we obtain a binary symmetric channel with
crossover probability $2p/(1+p)$.  
In the other case, where the measurement was $V_b$ and neither of the 
two possible states consistent with the first-stage code
is $b$, we gain no additional information, since 
both possible states remain equally likely.
The information gain from 
the second step is thus
\begin{equation}
{\textstyle \frac{1}{2}} (1+p) 
\left[1 -H\left(\frac{2p}{1+p}\right) \right] = 0.67673 \ \mathrm{bits}
\label{info-stage2}
\end{equation}
per signal.  
Adding the
information gains from the first two stages [Eqs.~(\ref{info-stage1})
and (\ref{info-stage2})] together gives 
\begin{equation}
\log_2 3 - H\left({\textstyle p; \frac{1-p}{2}; \frac{1-p}{2}}\right),
\end{equation}
which is the full information gain from
the measurement on the trine states that were lifted further out of the
plane; that this happens is, in some sense, an application of the chain rule 
for classical entropy~\cite{Cover}.

\newpage

\section{The upper bound on $C_{1,1}$}
\label{sec-exp-close}

In this section, we show that for the lifted trines $T(\alpha)$, if 
$\alpha$ is small, then the $C_{1,A}$ capacity
is exponentially close to the accessible information obtainable using
just two of our trines, showing that the three red curves in Fig.~\ref{fig-c1}
are exponentially close when $\alpha$ is close to 0.

First, we need to prove that in a classical channel,
changing the transition probabilities by
$\epsilon$ can only change the Shannon information by
$O(-\epsilon \log_2 \epsilon )$.
The Shannon capacity of a classical
channel with input distribution $p_i$ and
transition probabilities~$q_{ij}$ is the entropy of the output less
the entropy of the output given the input, or
\begin{equation}
I_S = -\sum_{j=0}^{N_{\rm out}-1}
\left( \sum_{i=0}^{N_{\rm in}-1} p_i q_{ij} \right)
\log_2 \sum_{i=0}^{N_{\rm in}-1} p_i q_{ij} 
+ \sum_{i=0}^{N_{\rm in}-1} \sum_{j=0}^{N_{\rm out-1}} p_i q_{ij} \log_2 q_{ij}.
\label{IS}
\end{equation}
Suppose we change all the $q_{ij}$ by at most $\epsilon$.  I claim that
the above expression changes by at most
$ -2 N_{\rm out} \, \epsilon \log_2 \epsilon$.
Each of the terms $q_{ij} \log_2 q_{ij} $ in the second term changes by at most
$-\epsilon\log_2 \epsilon$, and adding these changes (with weights $p_i$) gives 
a total change of at most $- N_{\rm out} \epsilon \log_2 \epsilon$.  
Similarly, each of the terms 
$\sum_i p_i q_{ij}$ in the first term of (\ref{IS})
changes by at most $\epsilon$, and there are at
most $N_{\rm out}$ of them, so we see easily that the first term 
also contributes 
at most $-N_{\rm out} \epsilon \log_2 \epsilon$ to the change.  
For $N_{\rm out} \leq 6$, which by the real version of Davies' theorem 
is sufficient for the optimum measurement on the lifted trines, we have that
the change is at most $-12 \epsilon \log_2 \epsilon$.

Next, we need to know that the $C_{1,1}$ capacity for the planar trines
is maximized using the probability distribution $(0,\frac{1}{2},\frac{1}{2})$, 
One can easily convince oneself of this by inspecting 
Figure~\ref{planartrines3d}.  We will discuss this at more length in Appendix A,
where we sketch a proof that the point $(0,\frac{1}{2},\frac{1}{2})$
is a local maximum for the accessible information.  The proof also shows that 
to achieve a capacity close 
to $C_{1,1}$ one must use a probability distribution and measurement close to
those achieving the optimal, a fact we will be using in this section.

Now, we can deal with lifted trines.  We will consider the trine states
$T(\alpha)$ for small~$\alpha$.   By moving each of the trines $T(\alpha)$ 
by an angle of 
$\phi = \arcsin \sqrt{\alpha} < 2 \sqrt{\alpha}$, we can obtain the 
planar trines $T(0)$.  
We now have that the difference between the transition probabilities for
$T(\alpha)$ and $T(0)$ for any rank 1 element in a POVM is at most $\phi$, 
since the transition probability for a fixed POVM element $r\proj{v}$
is a constant multiple (with 
the constant being $r\leq 1$) of the square of 
the cosine of the angle
between the vectors $\ket{v}$ and $\ket{T_b}$, and this angle changes by 
at most $\phi$.  Thus, by the lemma above, the $C_{1,1}$ capacity for
the lifted trines $T(\alpha)$ differs by no more than
$\delta = -12 {\phi} \log_2 {\phi}$ from the 
capacity obtained when the same probabilities (and measurement) are used 
for the planar trine states $T(0)$.

For the lifted trine states $T(\alpha)$, we know (from Section \ref{C11-sec})
that the $C_{1,1}$ capacity $C_{1,1}(\alpha)$ is greater than $C_{1,1}(0)$, 
the capacity for the planar trine states.  If we apply 
to the planar trine states $T(0)$ the same measurements and 
the same probability distribution that give the optimum $C_{1,1}$ capacity
for the lifted trine states $T(\alpha)$, we know that we have changed the 
capacity by at most $\delta = -12\phi \log_2 \phi$.  We thus have that 
$C_{1,1}(0) < C_{1,1}(\alpha) < C_{1,1}(0) + \delta$,  and that the
measurements and probability distribution that yield the optimum capacity
$C_{1,1}(\alpha)$ for the lifted trines $T(\alpha)$ must give a capacity
of at most $C_{1,1}(0) - \delta$ when applied to the planar trines.  This
limits the probability distribution and measurement giving the optimum 
capacity $C_{1,1}$ for the lifted trines.  For sufficiently small $\alpha$,
the optimum probability distribution on the trines must be 
close to $(0, \frac{1}{2},\frac{1}{2})$, and when the optimum projectors 
are projected onto
the plane, nearly all the mass must be contained in projectors within a
small angle of the optimum projectors for $C_{1,1}(0)$, namely
$\frac{1}{\sqrt{2}}(1, \pm 1)$.

We now consider the derivative in the information capacity obtained 
when the measurement is held fixed, and $p_0$ is increased at a rate
of 1 while $p_1$ and $p_2$ are decreased, each at 
a rate of $1/2$.  Taking the derivative
of (\ref{IS}), we obtain 
\begin{equation}
I_S'  = -\sum_{j=0}^{5}
\left( \sum_{i=0}^{2} p'_i q_{ij} \right)
\log_2 \sum_{i=0}^{2} p_i q_{ij} 
+ \sum_{i=0}^{2} \sum_{j=0}^{5} p'_i q_{ij} \log_2 q_{ij},
\label{ISprime}
\end{equation}
where $p_0' =1 $ and $p_1' = p_2' = -1/2$.  This derivative can be
broken into terms associated with each of the projectors in the measurement.
Namely, if the $j$th POVM element is ${r_j}\proj{v}$, then 
the associated term is 
\[
-r_j \left( \sum_{i=0}^{2} p'_i q_{iv} \right)
\log_2 \sum_{i=0}^{2} p_i q_{iv} 
+ r_j \sum_{i=0}^{2} p'_i q_{iv} \log_2 q_{iv},
\]
where $q_{iv} = |\braket{T_i}{v}|^2$, 
$p_0' =1$ and $p_1'=p_2' = -1/2$.  The $r_j$ can be factored out, and
this term can be written as $r_j I'_v$ where we define
\begin{equation}
I_v' =
- \left( \sum_{i=0}^{2} p'_i q_{iv} \right)
\log_2 \sum_{i=0}^{2} p_i q_{iv} 
+ \sum_{i=0}^{2} p'_i q_{iv} \log_2 q_{iv}.
\label{Ivprime}
\end{equation}
It is easy to see that for the planar trines, 
a projector $v_\theta = (\cos \theta, \sin \theta)$ with $\theta$ 
sufficiently close to $\pm\pi/4$, and a
probability distribution sufficiently close to $(0,\frac{1}{2},\frac{1}{2})$, 
the term $I'_{v_\theta}$ is approximately $-0.3227$ bits, the value of 
$I_{v_\theta}'$ at 
$\theta=\pm\pi/4$ and the probability distribution 
$(0,\frac{1}{2},\frac{1}{2})$.  
Similarly, for the planar trines, if the probability distribution is close to 
$(0, \frac{1}{2}, \frac{1}{2})$, 
$I_{v_\theta}'$ can never be much greater than $2$ bits, which is the 
maximum value for the probability distribution
$(0,\frac{1}{2}, \frac{1}{2})$ (occurring at $\theta = 0$).  
These facts show that if $\alpha$ 
is sufficiently small, then the formula (\ref{ISprime}) for the
derivative of the accessible 
information is negative and 
bounded above by (say) $-0.64$ bits when the planar trines are measured with 
the optimum POVM for $C_{1,1}(\alpha)$.  This is 
true because $\sum_{j} r_j =2$, 
and all but a $1-\epsilon$ fraction of the mass must be in
projectors $v_\theta$ for $\theta$ near $\pm \pi/4$; each of these projectors
will contribute at least $0.321r_j$ (say) to the derivative, and the projectors
with $\theta$ not near $\pm\pi/4$ cannot change this result by more than
$4 \epsilon$.   For the optimum
measurement and probability distribution for $C_{1,1}(\alpha)$ to have 
a non-zero value of~$p_0$ (from Section \ref{C11-sec} we know it does), 
this negative derivative must be
balanced by a positive derivative acquired by 
some projectors when the trines are lifted
out of the plane.  We will show that this can only happen when $p_0$ is
exponentially small in $1/\delta$; for larger values of $p_0$, the positive 
component acquired when the trines are lifted out of the plane is dwarfed
by the negative component retained from the planar trines. 

We have shown that $I_S'< -0.64$ bits near the probability distribution
$(0,\frac{1}{2},\frac{1}{2})$ when the optimal measurement 
for $T(\alpha)$ is applied to the planar trines,
assuming sufficiently small $\alpha$.   We also know 
from the concavity of the mutual information 
that for the lifted trines $T(\alpha)$ with $\alpha > 0$, the
derivative $I_S'$ is positive for any probability distribution 
$(p_0-t, p_1+t/2. p_2+t/2)$ where $(p_0, p_1, p_2)$ is the
optimal probability for $C_{1,1}$ capacity and $0 < t \leq p_0$.
Thus, we know 
that the negative derivative for the planar trines 
must be balanced by a positive derivative acquired by some projectors
when you consider the difference between the planar trines and the
lifted trines.  We will show that this can only happen when the probability
$p_0$ is exponentially small in the lifting angle $\phi$.  This shows that
at the probability distribution achieving $C_{1,1}$, $p_0$ is exponentially
small in $1/\phi = 1/\arcsin \sqrt{\alpha}$.

Consider the change in the derivative $I_{v_j}'$ for a
given projector $v_{j}$ when the trines $T(0)$ are lifted out of the plane
to become the trines $T(\alpha)$.  To make $I_{v_j}'$ positive,
this change must be at least 
$0.64$ bits.
Let the transition probabilities with the optimal 
measurement for $C_{1,1}(\alpha)$ be $r_j q_{iv_j}$ and the transition 
probabilities for the same measurement applied to the planar trines be
$r_j\tilde{q}_{iv_j}$.  Since the constant factors $r_j$ multiplying the
projectors sum to $3$, we have that the value $I_v'$ for one projector $v$
must change by at least $0.21$ bits, that is,
\begin{eqnarray*}
\nonumber
\left| 
\left( \sum_{i=0}^{2} p'_i \tilde{q}_{iv} \right)
\log_2 \sum_{i=0}^{2} p_i \tilde{q}_{iv} \right.
&-&\left( \sum_{i=0}^{2} p'_i q_{iv} \right)
\log_2 \sum_{i=0}^{2} p_i q_{iv} \\
&&\left. \quad + \sum_{i=0}^{2} p'_i \left( q_{iv} \log_2 q_{iv} 
- \tilde{q}_{iv}\log_2 \tilde{q}_{iv} \right)
\right| > 0.21
\label{Ivprimeterms}
\end{eqnarray*}
where $\tilde{q}_{iv} = |\braket{T_i(0)}{v}|^2$ and
${q}_{iv} = |\braket{T_i(\alpha)}{v}|^2$, as before. 
We know $|\tilde{q}_{iv} - q_{iv}| \leq \phi$.

We will first consider the last term of (\ref{Ivprimeterms}),
\[
\left|\sum_{i=0}^2 \sum_{j=0}^5 p'_i (q_{iv}\log_2 q_{iv} - \tilde{q}_{iv}
\log_2 \tilde{q}_{iv})\right|
\]
This is easily seen to be bounded by $-6 \phi \log_2 \phi$, which approaches
$0$ as $\alpha$ approaches $0$.

Next, consider the first terms of (\ref{Ivprimeterms}),
\begin{equation}
\left|
\left( \sum_{i=0}^{2} p'_i q_{iv}  \right)
\log_2 \sum_{i=0}^{2} p_i {q}_{iv} 
- \left( \sum_{i=0}^{2} p'_i \tilde{q}_{iv}  \right)
\log_2 \sum_{i=0}^{2} p_i \tilde{q}_{iv} 
\right|.
\label{secondterm}
\end{equation}
Bounding this is a little more complicated.  First, we will derive a 
relation among the values of $\tilde{q}_{iv}$ for different $i$. 
We use the fact that for the planar trines
\[
\ket{T_0(0)} = - \ket{T_1(0)} - \ket{T_2(0)}.
\]
Taking the inner product with $\bra{v_j}$, we get
\[
\braket{v_j}{T_0(0)} = - \braket{v_j}{T_1(0)} 
- \braket{v_j}{T_2(0)}.
\]
And now, using the fact that
$\tilde{q}_{iv} = |\braket{v_{j}}{T_i(0)}|^2$, we
see that
\begin{equation}
\tilde{q}_{0v} \leq 2 (\tilde{q}_{1v} +\tilde{q}_{2v} ).
\label{qrelation}
\end{equation}
Using (\ref{qrelation}), and the fact
that $p_1, p_2$ are close to $\frac{1}{2}$,
we have that
\begin{eqnarray}
\nonumber \sum_{i=0}^{2} p_i \tilde{q}_{iv} &\geq&  
\frac{3}{8}\tilde{q}_{1v} + \frac{3}{8}\tilde{q}_{2v} \\
&\geq&  \frac{1}{8}( \tilde{q}_{0v} +\tilde{q}_{1v} +\tilde{q}_{2v})
\label{two-to-three}
\end{eqnarray}
for sufficiently small $\alpha$.
We also need a relation among the $q_{iv}$.  We have
\begin{eqnarray}
\nonumber
q_{0v} + q_{1v} + q_{2v} &=&
\bra{v} \Big( \sum_{i=0}^2 \proj{T_i} \Big) \ket{v}\\
&\geq & \frac{1}{4}\phi^2,
\label{boundqsum}
\end{eqnarray}
where the second step follows because the minimum eigenvalue of
$\proj{T_0} +\proj{T_2} +\proj{T_2}$ is $\alpha > \phi^2/4$.  

We now are ready to bound the formula (\ref{Ivprimeterms}).  
We break it into two pieces; if this expression is at least 0.2 bits,
then one of these two pieces must be at least 0.1 bits.  
The two pieces are as follows:
\begin{equation}
\left|
\left( \sum_{i=0}^{2} p'_i (q_{iv}- \tilde{q}_{iv})   \right)
\log_2 \sum_{i=0}^{2} p_i {q}_{iv} 
\right|
\label{firstpiece}
\end{equation}
and
\begin{equation}
\left|
\left( \sum_{i=0}^{2} p'_i \tilde{q}_{iv}  \right)
\left( \log_2 \sum_{i=0}^{2} p_i {q}_{iv} -
\log_2 \sum_{i=0}^{2} p_i \tilde{q}_{iv} \right)
\right|.
\label{secondpiece}
\end{equation}

We first consider the case of (\ref{firstpiece}).  
Assume that it is larger than $0.1$ bits. Then
\begin{eqnarray}
\left|
\left( \sum_{i=0}^{2} p'_i (q_{iv}- \tilde{q}_{iv})   \right)
\log_2 \sum_{i=0}^{2} p_i {q}_{iv} 
\right| &\leq& 
- \left(\sum_{i=0}^2 |p_i'|  \phi \right) \log_2 \sum_{i=0}^2 p_0 q_{iv} \\
&\leq& - 2 \phi \log_2 \frac{p_0\phi^2}{4}
\end{eqnarray}
where the first step follows from the facts 
that $|q_{iv} - \tilde{q}_{iv}| < \phi$
and that $p_0$ is the smallest of the $p_i$, and 
the second step follows from (\ref{boundqsum}).
Thus, if the quantity (\ref{firstpiece}) is at 
at least $0.1$, we have that
\[
p_0 < \frac{4}{\phi^2} 2^{- 0.05 / \phi}
\]
showing that $p_0$ is exponentially small in $1/\phi$.

We next consider the case (\ref{secondpiece}).  Assume that 
\[
\left|
\left( \sum_{i=0}^{2} p'_i \tilde{q}_{iv}  \right)
\left( \log_2 \sum_{i=0}^{2} p_i {q}_{iv} -
\log_2 \sum_{i=0}^{2} p_i \tilde{q}_{iv} \right)
\right|
\]
is larger than $0.1$ bits.  We know that 
\[
\left|\sum_{i=0}^2 p'_i \tilde{q}_{iv}\right|  \leq \sum_{i=0}^2  |p'_i| =2
\]
Thus, for (\ref{secondpiece}) to be larger than $0.1$, we must have that
\[
\left| \log_2 \frac{\sum_{i=0}^2 p_i q_{iv}}{\sum_{i=0}^2 p_i \tilde{q}_{iv}}
\right| > 0.05\,.
\]
We know that
the numerator and denominator inside the logarithm differ by at most~$\phi$.
It is easy to check that if $|\log_2(x/y)| > 0.05$, and $x-y \leq \phi$,
then both $x$ and $y$ are at most $15 \phi$.  Thus,
\begin{equation}
\label{boundof15}
\sum_{i=0}^2 p_i \tilde{q}_{iv} < 15 \phi.
\end{equation}
Further, 
\begin{eqnarray}
\label{one-twenty}
\left|\sum_{i=0}^2 p'_i \tilde{q}_{iv}\right| &\leq& 
\sum_{i=0}^2 \tilde{q}_{iv} \\
&\leq& 8\sum_{i=0}^2 p_i \tilde{q}_{iv}\nonumber\\
&\leq& 120\phi.\nonumber
\end{eqnarray}
where the second inequality follows by (\ref{two-to-three}) and the third
by (\ref{boundof15}). 

Since the two terms in (\ref{secondpiece}) are
of opposite signs, if they add up to at least 0.1 bits, at least one of them 
must exceed $0.1$ bits by itself.  We will treat these two cases separately.  
First, assume that the first term exceeds $0.1$.  Then
\begin{eqnarray*}
0.1 &\leq &
-\left|\sum_{i=0}^2 p_i' \tilde{q}_{iv} \right| 
\log_2 \sum_{i=0}^2 p_i q_{iv}\\
&\leq & -\left|\sum_{i=0}^2 p_i' \tilde{q}_{iv} \right| 
\log_2 \sum_{i=0}^2 p_0 q_{iv}\\
& \leq & 
-120 \phi \log_2 \frac{p_0 \phi^2}{4},
\end{eqnarray*}
where the last inequality follows from (\ref{boundqsum}) and 
(\ref{one-twenty}).
If this is at least $0.1$, then we again have that $p_0$ is exponentially
small in $1/\phi$.

Finally, we consider the case of the term 
\[
-\left|\sum_{i=0}^2 p_i' \tilde{q}_{iv} \right| 
\log_2 \sum_{i=0}^2 p_i \tilde{q}_{iv}.
\]
We have by (\ref{two-to-three}) that
\[
-\left|\sum_{i=0}^2 p_i' \tilde{q}_{iv} \right| 
\log_2 \sum_{i=0}^2 p_i \tilde{q}_{iv}  \leq 
- \left( \sum_{i=0}^2 \tilde{q}_{iv}\right)
\log_2 \frac{1}{8} \sum_{i=0}^2  \tilde{q}_{iv},
\]
which, since $\sum_{i=0}^2 \tilde{q}_{iv} < 120 \phi$, can never exceed $0.1$
for small $\phi$, as it is of the form $-8x \log x$ for a small $x$.

Since $p_0$ is exponentially small in $1/\phi$, we have that the difference
between the $C_{1,1}$ capacity using only two trines and that using all three
trines is exponentially small in $\frac{1}{\sqrt{\alpha}}$, showing that
our guess in Section~\ref{C11-sec} are exponentially close to
the correct $C_{1,1}$ capacity as $\alpha$ goes to $0$, and thus
showing that $C_{1,A}$ is strictly larger than 
$C_{1,1}$ in a region near $\alpha=0$.  

\section{$C_{1,1}=C_{1,A}$ for two pure states}
\label{sec-upperbound}

In this section, we prove that for two pure states, $C_{1,1} = C_{1,A}$. 
We do this by giving a general upper bound on $C_{1,A}$ based on a 
tree construction, 
We then use the fact that for two pure states, accessible information
is concave in the probabilities
of the states (proved in Appendix B) to show that
this upper bound is equal to $C_{1,1}$
for ensembles containing only two pure states.  

For the upper bound, we consider a class of trees, with 
an ensemble of states associated with each node.  The action of Bob's 
measurement protocol on a specific signal will generate such
a tree, and analyzing this tree will bound the amount of information Bob
can on average extract from that signal.  Associated
with each tree will be a capacity, and the supremum over all trees will
give an upper bound for $C_{1,A}$.  

We now describe our tree construction in general.  
Let us suppose that Alice can convey to Bob one of $m$ possible signal 
states.  Let these states be $\rho_i$, where $1 \leq i \leq m$.
To each tree node we assign $m$ density matrices and $m$ associated
probabilities (these will not be normalized, and so may sum to less than 1).  
For node ${\bf x}$ of the tree, we associate some POVM element $E_{\bf x}$,
and the $m$ density matrices $E_{\bf x}^{1/2} 
\rho_i E_{\bf x}^{1/2}/\Tr E_{\bf x} \rho_i$, 
where $\rho_i$ are the original signal states.  
(We may omit the normalization factor of
$\Tr E_{\bf x}\rho_i$ in this discussion when it is clear from context.)
For the root node ${\bf r}$, 
the POVM element $E_{\bf r}$ is the identity matrix $I$, and   
the probability $p_{{\bf r},i}$ is the probability that this
signal is $\rho_i$.
A probability $p_{\bf x}$ can be associated with node ${\bf x}$ by summing 
$p_{\bf x} = \sum_{i=1}^m p_{{\bf x},i}$.  For the root,
$p_{\bf r} = 1$.  For any node ${\bf x}$, 
its associated probability
$p_{\bf x}$ will be equal 
to the sum of the probabilities $p_{{\bf y}_j}$ associated with 
its children ${\bf y}_j$.
There are two classes of nodes, distinguished by the means of obtaining
its children from 
the node.  The first class we call measurement nodes and the second we
call probability refinement (or refinement) nodes.
For a measurement node $\bf x$, we assign to
each of the children ${\bf y}_j$ a POVM elements 
$E_{{\bf y}_j}$, where  $\sum_j E_{{\bf y}_j} = E_{\bf x}$.  
The density matrices associated with a child of ${\bf x}$ will be
$E_{{\bf y}_j}^{1/2} \rho_i E_{{\bf y}_j}^{1/2}/\Tr E_{{\bf y}_j} \rho_i$, and
the probability associated
with the density matrix 
$E_{{\bf y}_j}^{1/2} \rho_i E_{{\bf y}_j}^{1/2}/\Tr E_{{\bf y}_j} \rho_i$ 
will be
$p_{{\bf y}_j,i} = p_{{\bf x},i} 
\Tr (E_{{\bf y}_j} \rho_i) / \Tr (E_{\bf x} \rho_i)$.  
Finally, we define the information gain associated with a node ${\bf x}$.
This is $0$ for nodes which are not measurement nodes, and
\[
I_{\bf x} = p_{\bf x} 
H\left(\left\{\frac{p_{{\bf x},i}}{p_{\bf x}}\right\}\right) - 
\sum_k p_{{\bf y}_k} 
H\left(\left\{\frac{p_{{\bf y}_k,i}}{p_{{\bf y}_k}}\right\}\right),
\]
where $H(\{q_i\})$ is the Shannon information $\sum_i q_i \log_2 q_i$
of the probability 
distribution $\{q_i\}$.  

We now explain why we chose this formula.
We consider applying a measurement to the ensemble associated
with node ${\bf x}$.  
This ensemble contains the state
$E_{\bf x}^{1/2} \rho_i E_{\bf x}^{1/2}/\Tr (\rho_i E_{\bf x})$ 
with probability $p_{{\bf x}, i}/p_{\bf x}$.  
Let us apply the measurement that takes
$\rho$ to $A_k \rho A_k^{\dag}$ with probability 
$\Tr A_k^\dag A_k \rho_i$, where $\sum_k A_k^\dag A_k = I$.  
Each child ${\bf y}_k$ of ${\bf x}$ is associated with one of the
matrices $A_k$.  
Let $E_{{\bf y}_k} = E_{\bf x}^{1/2} A_k^\dag A_k E_{\bf x}^{1/2}$.  
Then $\sum_k E_{{\bf y}_k} = E_{\bf x}$.  
Now, after we apply $A_k$ to $E_{\bf x}^{1/2} \rho_i E_{\bf x}^{1/2}$, we 
obtain the state
$A_k E_{\bf x} \rho_i E_{\bf x} A_k^\dag$.  This happens with probability 
\[
\frac{\Tr A_k E_{\bf x}^{1/2} \rho_i E_{\bf x}^{1/2} A_k^{\dag}}
{\Tr  E_{\bf x} \rho_i }
= \frac{\Tr E_{{\bf y}_k}\rho_i}{\Tr E_{\bf x}\rho_i}.  
\]
The state we obtain,
$A_k E_{\bf x}^{1/2} \rho_i E_{\bf x}^{1/2} A_k^\dag$, 
is unitarily equivalent
to $E_{{\bf y}_k}^{1/2} \rho_i E_{{\bf y}_k}^{1/2}$, so this 
latter state can be
obtained by an equivalent measurement.  
The information $I_{\bf x}$ associated with the node $\bf x$ is the 
probability 
of reaching the node times the Shannon information gained by 
this measurement if the node is reached.   Summing $I_{\bf x}$ over 
all the nodes ${\bf x}$ of the tree gives the expected information gain
by measurement steps.   

The second class of nodes are probability refinement nodes
(which we often shorten to refinement nodes).  
Here, for all the children  $\{ {\bf y}_k\}$ of ${\bf x}$,
$E_{\bf x}=E_{{\bf y}_k}$.  We assign probabilities $p_{{\bf y}_k,i}$ 
to the children ${\bf y}_k$ so that
$\sum_k p_{{\bf y}_k,i} = p_{{\bf x},i}$.  For this class of nodes,
we define $I_{\bf x}$ to be $0$.  These nodes correspond to steps in the
protocol where additional information is gained about one signal state
in the codeword by measuring different signal states in the codeword.

To find the upper bound on the $C_{1,1}$ capacity for a set of
states $\{\rho_i\}$, we take the supremum over the information gain
associated with all trees of the above form.
That is, we try to maximize $\sum_{\bf x} I_{\bf x}$ over all 
probability distributions
$p_{{\bf r},i}$ on the root node ${\bf r}$, all ways of splitting 
$E_{\bf x} = \sum_k E_{{\bf y}_k}$ for measurement nodes ${\bf x}$,
and all ways of splitting probabilities 
$p_{{\bf x},i} = \sum_k p_{{\bf y}_k,i}$ for refinement nodes
${\bf x}$ and signal states $i$.  (And if this maximum is not attained,
we take the supremum.)

To prove this upper bound, we
track the information obtained from a single
signal (i.e., channel output) $S_\nu$ in the 
protocol used by Alice and Bob.  We assume that
Alice sends Bob a set of states, and Bob performs measurements on 
them one at a time.  We keep track at all times $t$ (i.e., for all
nodes {\bf x} of the tree) of the probability
that signal $S_\nu$ is in state $E_{\bf x}^{1/2} \rho_{i}E_{\bf x}^{1/2}$.  
There are two cases, depending
on which signal Bob measures.  
In the first case, when Bob measures signal $S_\nu$, we perform
a measurement on the current tree node $\bf x$ that splits each
of the possible values of $\rho_{t,i}$ for this signal $S_\nu$
into several different
values.  This case corresponds to a measurement node of the tree.
We can assume without loss of generality that 
for his measurement Bob uses the canonical type of operators 
discussed above,
so that $E_{\bf x}^{1/2}\rho_i E_{\bf x}^{1/2}$ goes to 
$E_{{\bf y}_k}^{1/2}\rho_iE_{{\bf y}_k}^{1/2}$ 
with probability $\Tr E_{{\bf y}_k} \rho_i/ \Tr E_{\bf x} \rho_i$.  
Thus, we
now have several different ensembles of density matrices, the $k$th of which 
contains 
$E_{{\bf y}_k}^{1/2}\rho_iE_{{\bf y}_k}^{1/2}/\Tr E_{{\bf y}_k} \rho_i$ 
with (unnormalized) probability 
$p_{{\bf x},i} \Tr E_{{\bf y}_k} 
\rho_i/ \Tr E_{{\bf x}} \rho_i = p_{{\bf y}_k,i}$.
In this step Bob can extract some information
about the original codeword, and 
the amount of this information is at most $I_{\bf x}$.

The other case comes when Bob measures signals
than $S_\nu$.  These steps can provide additional information about 
the signal $S_\nu$, so 
if the probability distribution before this step contained 
$E_{\bf x}^{1/2} \rho_iE_{\bf x}^{1/2} $ with 
probability $p_{{\bf x},i}$, we
now have several distributions, each assigned to a child of ${\bf x}$; 
the $j$th distribution contains
$E_{\bf x}^{1/2} \rho_i E_{\bf x}^{1/2}$ with (unnormalized) probability 
$p_{{\bf y}_j,i}$  Here, we must have 
$\sum_j p_{{\bf y}_j,i}= p_{{\bf x},i}$.   This kind of step
corresponds to a
probability refinement node in the tree.  The information gained by these
measurement steps can
be attributed to the signals that are actually measured
in these steps, so we need not attach any information gain to
the refinement steps in the
tree formulation.  Averaging the information
gain over the trees associated with all the signals
gives the capacity of the protocol, which is the expected information
gain per signal sent. 

There are several simplifying assumptions we can make about the trees.  First,
we can assume that nodes just above leaves are measurement nodes
that contain only rank 1 projectors, since any refinement node having no
measurement nodes below it can be eliminated without reducing the 
information content of the tree, and since the last measurement
might as well extract as much information as possible. 
We could assume that the types of the nodes are alternating, since 
two nodes of the same type, one a child of the other, 
can be collapsed into one node.  In the sequel, we will perform this
collapse on the measurement nodes, so we assume that all the children
of measurement nodes are refinement nodes.  We could also
(but not simultaneously) assume that every node has degree two, since 
any measurement with more than two outcomes can be replaced with an
equivalent sequence of measurements, each having only two outcomes,
and any split in probabilities
can be replaced by an equivalent sequence of splits.  In the sequel
we will assume that all the probability refinement nodes are of degree
two.

One interesting question is whether any tree of this form has an 
associated protocol.  The upper bound will hold whether or not this
is the case, but if there are trees with no associated protocols, 
the bound may not be tight.  We do not know the answer to this, but suspect
that there are trees with no associated protocols.  Our (vague) intuition 
is that if the root node is a measurement node with no associated
information gain, and all of the children of this node are refinement
nodes, there appears to be no way to obtain the information needed to 
perform one of these refinement steps without also obtaining the 
information needed to perform the all the other refinement steps 
of the root node.  However, making this much information available at
the top node would reduce the information that could be obtained using later
measurement nodes.   It is possible that this
difficulty can be overcome if there is a feedback
channel available from the receiver to the sender.  We thus boldly
conjecture
\begin{conjecture}
If arbitrary use of a classical feedback channel from the sender to 
the receiver is available for free, then the adaptive capacity with 
feedback $C_{1,AF}$ is given by the supremum over all trees of the 
above type of the information associated with that tree.
\end{conjecture}

As mentioned above, the supremum of the extractable information over all trees 
is an upper bound on $C_{1,A}$, 
since it is at least as large as the information corresponding to any 
possible adaptive protocol.  We now restrict our discussion to the case 
of ensembles consisting of two pure states, and prove that in this case
we have equality, since both of these bounds are equal to the $C_{1,1}$
capacity.  Consider a tree which gives
a good information gain for this ensemble (we would say maximum, but 
have no proof that the supremum is obtainable).
There must be a deepest refinement node, so all of its descendents 
are measurement nodes.  We may without
loss of generality assume that this deepest refinement node has only
two children.  Each of these two children has an associated
ensemble consisting of
two pure states with some probabilities.  The maximum 
information tree will clearly assign the optimum measurement to these
nodes.  However, an explicit expression for this optimum measurement is
known \cite{fuchs-thesis,Levitin-ai,Levitin-conj,Osaki-ai}, 
and as is proved in Appendix B, the accessible information for
ensembles of two given pure states is concave 
in the probabilities of the states.  Thus,
if we replace this refinement node with 
a measurement node, we obtain a tree with a higher associated information 
value.  Using induction, we can perform a series of such steps which
do not decrease the information gain associated with the
tree while collapsing everything to a single measurement.  Thus, for
two pure states, we have $C_{1,1} = C_{1,A}$.

The above argument would work to show that $C_{1,1} = C_{1,A}$ for 
ensembles consisting of two arbitrary density matrices if we could show
that the accessible information for two arbitrary density matrices is
concave in the probabilities of these two density matrices.  
It would seem intuitively that this should be true, but we 
have not been
able to prove it.  It may be related to the 
conjecture~\cite{Levitin-conj,Fuchs}
that the optimal accessible information for two arbitrary density 
matrices can always be achieved 
by a von Neumann
measurement.
This has been proved in two dimensions \cite{Levitin-conj}, and is supported
by numerical studies in higher dimensions \cite{Fuchs}.
We thus conjecture:
\begin{conjecture}
$C_{1,A} = C_{1,1}$ for two mixed states in arbitrary dimensions.  
\end{conjecture}

In fact, the proof in this section will work for any upper bound on accessible
information which has
both the concavity property and the property that if a measurement is made
on the ensemble, the sum of 
the information extracted by this measurement and
the expected upper bound for the resulting 
ensemble
is at most the original upper bound.  The Fuchs-Caves 
bound~\cite{fuchs-caves-nyas}
(which was Holevo's original bound) may have these properties; we have
done some numerical tests and have not found a counterexample.
For 3 planar trine states with equal probabilities, this gives an upper 
bound of approximately 0.96.

\begin{figure}[tbp]
\epsfxsize=\textwidth
\begin{center}
\phantom{,}\epsfig{width=\textwidth,file=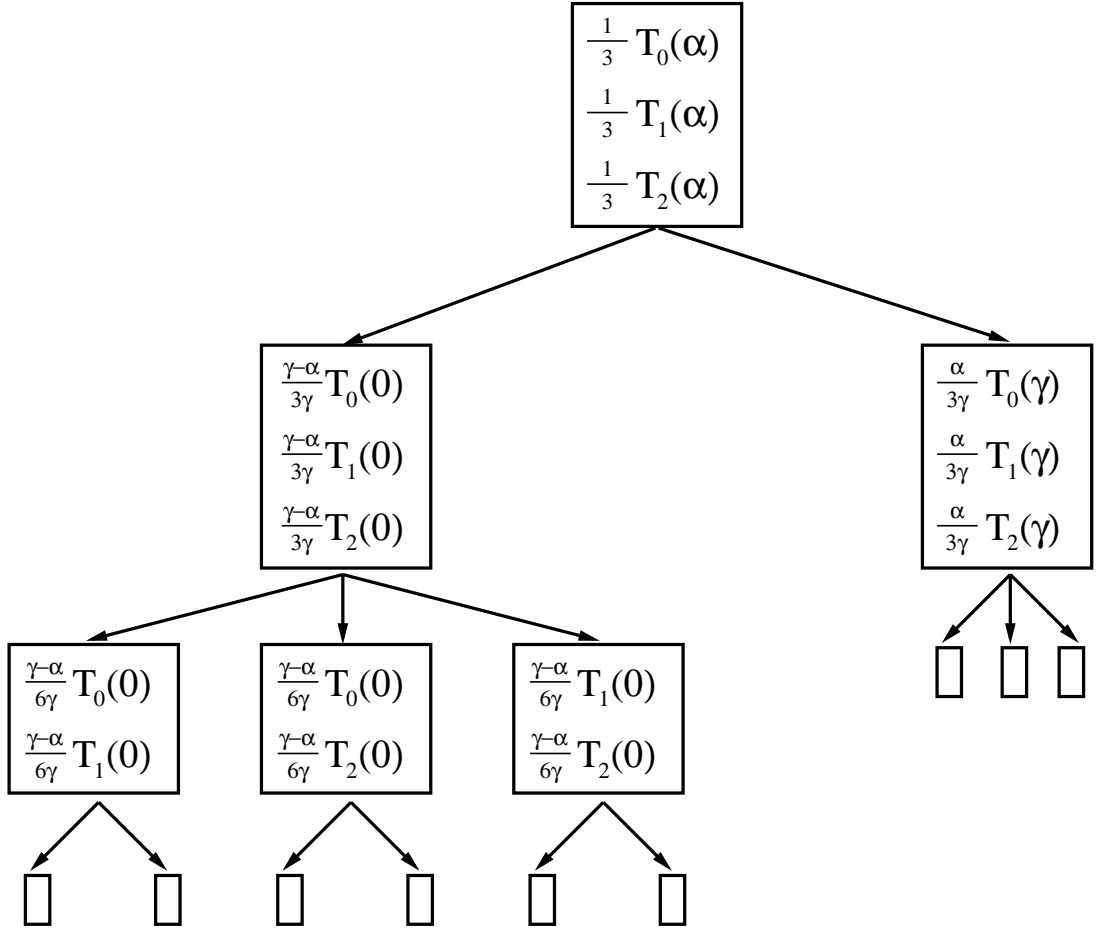}
\end{center}
\caption{\small 
The tree corresponding to the best adaptive protocol 
of Section~\protect{\ref{sec-adaptive}}.  To simplify the 
diagram, we do not give 
the probabilities and states in the ensembles of the leaves of this tree,
which are represented here by empty boxes.  
Since they are reached by the final measurement, which projects onto a 
rank 1 density matrix, the quantum states
corresponding to these nodes are
now completely reduced, and no further information can be extracted
from these ensembles.
The probabilities can be computed from the discussion in
Section~\protect{\ref{C11-sec}}.
\label{exampletree}
}
\end{figure}

\newpage

\section{Discussion}
\label{sec-discussion}

If we force Bob to measure his signals sequentially, so that he must complete
his measurement on signal $k$ before he starts measuring signal $k+1$
(even if he can adaptively choose the order he measures the signals in and even
if a feedback channel is applied from Bob to Alice), Bob
can never achieve a  capacity greater than $C_{1,1}$.  This can easily
be seen.  Without decreasing the capacity, we assume that Bob uses a 
feedback channel to send all the information that he has back    
to Alice.  This information consists of the results of the measurement
and the measurement that he plans to perform next.  
The ensemble of signals that Alice now sends Bob can convey no more
information than the optimal set of signals for this measurement.  
However, it now follows from 
classical information theory that such a protocol can never have a
capacity greater than the sum of the optimal information gains for
all these measurements, which is at most $C_{1,1}$.  

It is thus clear that the advantage of adaptive protocols is obtained 
from the fact that Bob can adjust subsequent measurements of a signal 
depending on the outcome of the first round of measurements on the entire 
codeword.
In the information 
decision tree of Section \ref{sec-upperbound} for our protocol 
(see Figure \ref{exampletree}).
the crucial fact is that we first either project each of the trine states 
into the plane or lift it up.  We then arrange to distinguish between 
only two possible states for those signals that were projected into the 
plane, and among all three possible states for those signals that were lifted. 

As we showed in Section~\ref{sec-upperbound}, $C_{1,1} = C_{1,A}$ for
two pure states, and this proof can be extended to apply to
two arbitrary states if a very plausible conjecture on the accessible 
information for a two-state ensemble holds.   For three states, even in 
two dimensions, the same upper bound proof cannot apply.  However, for
three states in two dimensions, it may still be that $C_{1,1} = C_{1,A}$.
We have unsuccessfully tried to find strategies that perform better 
than the $C_{1,1}$ capacity for the three planar trine states, and we now
suspect that the adaptive capacity is the same as the $C_{1,1}$ capacity
in this case, and that this is also the case for arbitrary
sets of pure states in two dimensions.  
\begin{conjecture}
For an arbitrary set of pure states in two dimensions, $C_{1,1} = C_{1,A}$,
and in fact, this capacity is achievable by using as signal states the two
pure states in the ensemble with inner product closest to 0.
\end{conjecture}

For general situations, we know very little about $C_{1,A}$. 
In fact, we have no
good criterion for deciding whether $C_{1,A}$ is strictly
greater than $C_{1,1}$.
Another question is whether entangled inputs could improve the adaptive 
capacity.  That is, whether $C_{1,A} = C_{\infty, A}$, where $C_{\infty,A}$
is the capacity given entangled inputs and single-signal, but adaptive,
measurements.  

\section*{Acknowledgments}
I would like to thank David Applegate for the substantial help he gave 
me with running CPLEX in order to compute the graphs in Appendix C, 
Chris Fuchs for valuable discussion on accessible information and $C_{1,1}$ 
capacity, and John Smolin for running computer searches looking for 
the $C_{1,1}$ capacity for the lifted trine states.
I would also like to thank Vinay Vaishampayan for inviting me to give
the talk on quantum information theory which eventually gave
rise to this paper.

\newpage
\section*{Appendix A: The $C_{1,1}$ capacity for the planar trines}

Next, we discuss the $C_{1,1}$ capacity for trines in the plane.  For
Section \ref{C11-sec}, we needed to show two things.  First, that $C_{1,1}$ 
for the planar trines was maximized using the probability distribution 
$\Pi_2 = (0,\frac{1}{2},\frac{1}{2})$, and second,
that any protocol with capacity close to $C_{1,1}$ must use 
nearly the same probability distribution and measurement
as the optimum protocol achieving $C_{1,1}$.
We show that in the neighborhood of the probability
distribution $\Pi_2$, the optimum measurement for accessible
information contains only two projectors,
From this proof, both facts can be easily deduced; we provide a proof
of the first, the second follows easily from an examination of our proof.

We first show that if an optimum measurement for accessible information
has only $k$ projectors, then at most $k$ different input states are
needed to achieve optimality.  This result is a know classical result; 
for completeness, I provide a brief proof.
Shannon's formula for
the capacity of a classical channel is the
entropy of the average output less the average entropy of the output.
It follows that the number of input states of a classical
channel needed to achieve optimality never exceeds the number of output
states.
If there are $k$ outcomes, and $k' > k$ input states, then the 
output probability distribution can be held fixed on a 
($k'-k$)-dimensional subspace of
the input probability distributions.  By the linearity of the
average entropy, the minimum average entropy can be achieved at a point
of that subspace which has only $k$ non-zero probabilities on the input
states.  Thus, if the optimal measurement is a von Neumann measurement, 
only two trines are required to achieve optimality.  

We associate to each projector $v_\theta = (\cos \theta, \sin \theta)$ 
an information quantity depending on the probability distribution
$\Pi = (p_0, p_1, p_2)$, namely
\[
I_\Pi(\theta) =
- \left(\sum_{i=0}^2 p_i q_{i,\theta} \right) 
\log_2 \sum_{i=0}^2 p_i q_{i,\theta}
+ \sum_{i=0}^2 p_i q_{i,\theta} \log_2 q_{i,\theta} 
\]
where $q_{i,\theta} = |\braket{T_i}{v_\theta}|^2$.
The accessible information for a measurement using POVM elements 
$r_j \proj{v_{\theta_j}}$ is $\sum_j r_j I_\Pi(\theta_j)$.
Now, we need to find the projectors that form a POVM, 
and maximize the accessible information.  
If we have projectors $v_{\theta_i}$ with associated weights 
$r_i$,
the constraints that the projectors form a POVM are:
\begin{eqnarray}
\sum_i r_i \cos^2 \theta_i &=& 1\\
\sum_i r_i \sin^2 \theta_i &=& 1\\
\sum_i r_i \sin \theta_i \cos \theta_i &=& 0.
\end{eqnarray}
These constraints are equivalent to
\begin{eqnarray}
\label{first-constraint}
\sum_i p_i &=& 2\\
\sum_i p_i \cos 2\theta_i &=& 0\\
\sum_i p_i \sin 2\theta_i &=& 0.
\label{last-constraint}
\end{eqnarray}
We wish to find projectors such that $\sum_i r_i I_\Pi(\theta_i)$ is maximum,
given the linear
constraints (\ref{first-constraint}--\ref{last-constraint}).
This is a linear programming problem.  
The duality theorem of linear programming says that this maximum is equal to
the twice the minimum $\alpha$ for which there is a $\sigma$ and a $\beta$ 
such that the inequality
\begin{equation}
\alpha + \beta \sin ( 2\theta + \sigma) \geq I_\Pi(\theta)
\label{lpdual}
\end{equation}
holds for all $\theta$.  (The factor of 2 comes from the right hand side of
Eq.~(\ref{first-constraint}).) It is easy to see that the
sine function of (\ref{lpdual}) and the function $I_\Pi(\theta)$ are either 
tangent at two values of $\theta$ differing by $\pi/2$, or 
are tangent at three values of $\theta$ (or more, in degenerate cases),
as otherwise a different sine function with a smaller $\alpha$ would
exist.  
If they are tangent at two points, then the optimal measurement is 
a von Neumann measurement, as it contains only two orthogonal projectors.

\begin{figure}[tbh]
\vspace*{1.6in}
\begin{center}
\epsfig{figure=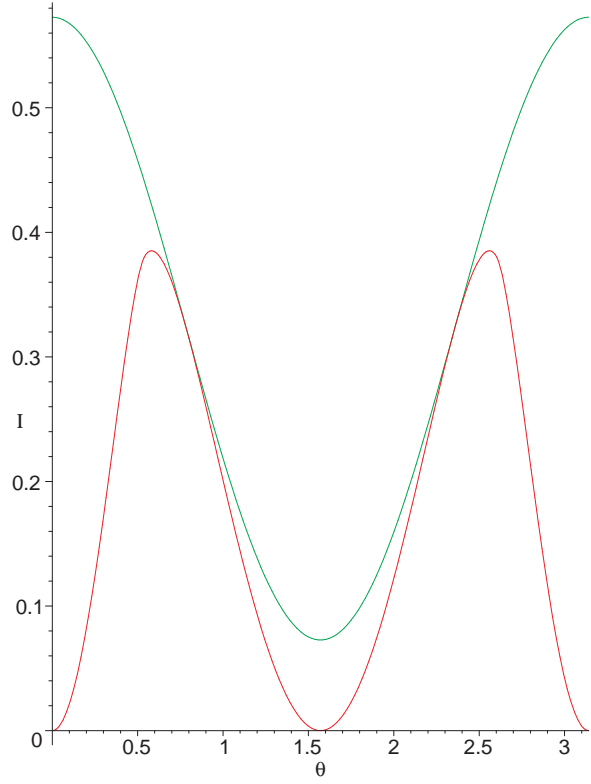,height=3in}
\end{center}
\caption{ \small The red curve is $I(\theta)$ for the probability
distribution $(0,\frac{1}{2},\frac{1}{2})$.  The green curve is
$\frac{1}{2}(1-H(\frac{1}{2} + \frac{\sqrt{3}}{4})) + \frac{1}{4}
\cos 2 \theta$.  These curves are tangent at the points $\pi/4$ and
$3\pi/4$, showing that the optimum measurement for accessible information
is the von Neumann measurement with projectors $v_{\pi/4}$ and 
$v_{3\pi/4}$.  It yields 
$1-H(\frac{1}{2} + \frac{\sqrt{3}}{4})$ bits of accessible information. 
\label{sine-I-graphs}
}
\end{figure}

For the probability distribution $\Pi_2 = (0, \frac{1}{2},\frac{1}{2})$, 
the two functions $I_{\Pi_2}(\theta)$ and 
\begin{equation}
\frac{1}{2}
\left(1-H\left(\frac{1}{2}-\frac{\sqrt{3}}{4}\right)\right)- 
\frac{1}{4} \sin(2\theta - \pi/2)
\label{optsinecurve}
\end{equation}
are plotted in Figure \ref{sine-I-graphs}.  
One can see that 
the sine function is greater than the function $I(\theta)$, and the functions are
tangent at the two points $\theta = \pi/4$ and $\theta=3\pi/4$, which 
differ by $\pi/2$.  
Hence, the linear program has an optimum of
$H(1/2 + \sqrt{3}/4)= 0.35458$ bits, and 
the optimal
measurement is a von Neumann measurement with projectors $v_{\pi/4}$
and $v_{3\pi/4}$,
and yielding $1-H(1/2+\sqrt{3}/4) = 0.64542$ bits of accessible information.

We wish to show that for all probability distributions $\Pi$ near $\Pi_2$,
the two functions behave similarly to the way they behave in 
Figure~\ref{sine-I-graphs}.
As the details of this calculation are involved and not particularly
illuminating, we leave them out, and merely sketch the outline
of the proof.

The first step is to show that for any function $I_\Pi(\theta)$ 
obtained using a
probability distribution $\Pi$ close to $(0,\frac{1}{2},\frac{1}{2}$),
there is a sine function close to the original sine
function (\ref{optsinecurve})
which always exceeds $I_{\Pi}(\theta)$ and is tangent to $I_\Pi(\theta)$
at two points in regions near $\theta= \pi/4$ and $\theta=3\pi/4$. 
We do this by finding
values $\theta_1$ and $\theta_2$
in these regions which differ by $\pi/2$ and such that the derivative 
$I'_\Pi(\theta) = dI_\Pi(\theta)/d\theta$ evaluated at $\theta_1$ and 
$\theta_2$ has
equal absolute values but opposite signs; these two points
define the sine function.  We show that these two points must exist by finding 
an $\epsilon$ such that 
\begin{eqnarray*}
I'_\Pi(\pi/4 - \epsilon) + I'_\Pi(3\pi/4 - \epsilon) &>& 0,  
\ \ \ \ \ \ \mathrm{and}\\
I'_\Pi(\pi/4 - \epsilon) + I'_\Pi(3\pi/4 - \epsilon) &<& 0,
\end{eqnarray*}
and using the continuity of the first derivative of $I_{\Pi}(\theta)$.
This $\epsilon$ is calculated by using the fact that if 
the probability distribution $\Pi$ is close to $\Pi_2$, then $I'_\Pi$ 
is close to $I'_{\Pi_2}$.

To show that  $v_{\theta_1}$ and $v_{\theta_2}$ are indeed the optimal 
projectors for the probability distribution $\Pi$, we need to show that
except at the points $\theta_1$ and $\theta_2$, the sine function we have
found is
always greater than $I_\Pi(\theta)$.  We do this in two steps.
First, we show that the sine function is greater than $I_\Pi(\theta)$
in the regions far from the points of tangency.  This can be done 
using fairly straightforward estimation techniques, since outside 
of two regions centered around the values
$\theta = \pi/4$ and $\theta = 3\pi/4$, these functions do not approach 
each other closely. 
Second, we show that the second derivative of the sine function is 
strictly greater than the second derivative $d^2I_\Pi(\theta)/d \theta^2$ 
in the two regions near the points of tangency.  This shows that the 
function $I_\Pi(\theta)$
cannot meet the sine function in more than one point in each 
of these regions.  

Our calculations show that for probability distributions
within $0.001$ of 
$(0,\frac{1}{2}, \frac{1}{2})$ in the $L_1$ norm,
there are only two 
points of tangency. Recall the fact that to achieve optimal capacity,
classical channels never require more input states than
output states.  This shows that
using the same measurement and adjusting the smallest probability in $\Pi$
to be $0$ will improve the accessible information, showing
that this accessible information is at most that achievable using only
two trines, namely $1-H(1/2+\sqrt{3}/4) = 0.64542$.
We need now only show that for points outside this
region, the accessible information cannot approach 0.64542; while we have
not done this rigorously, 
the graph of Figure \ref{planartrines3d}, and similar graphs 
showing in more detail the
regions near the points of tangency, are extremely strong evidence that 
this is indeed the case.  In fact, numerical experiments appear to 
show that if the minimum 
probability 
of a trine state is less than $0.06499$, then there are only two 
projectors in the optimal measurement; the probability distribution 
containing the minimum probability and 
requiring three projectors is approximately
$(0.065, 0.4675, 0.4675)$.

\newpage
\section*{Appendix B: Convexity of accessible information on two pure states}

For section \ref{sec-upperbound}, we needed a proof that the accessible 
information on two pure states $v_1$ and $v_2$ is a concave function in 
the probabilities of these pure states.  As opposed to the rest of the
paper, all logarithms in this section will have base e.

We first prove an inequality that will be used later.
For $0 \leq x < 1$, 
\begin{equation}
F(x) = \frac{2x}{1-x^2} - \log\left(\frac{1+x}{1-x}\right) \geq 0
\label{eq-inequality}
\end{equation}
It is easy to see that for $x=0$, both terms are 0.  Differentiating
and simplifying, we get
\[
F'(x) = \frac{4x^2}{(1-x^2)^2}
\]
which is positive for $0 \leq x < 1$, so $F(x) \geq 0$ in this range.

We now prove that the accessible information is a concave function in $p$
for an ensemble consisting of two pure states, $\ket{v_1}$ with probability
$p$ and $\ket{v_2}$ with probability $1-p$.
The formula for this accessible information is 
\[
I_{acc} = H(p) - H\left({\textstyle \frac{1}{2} 
+ \frac{1}{2}\sqrt{1-4 \kappa p (1-p)}}\right) 
\]
where $\kappa = | \braket{v_1}{v_2}| ^2$,
and $H$ is the Shannon entropy function (which we will take to the base $e$
in this section).  Proofs of this formula can be found in 
\cite{Levitin-ai,fuchs-thesis,Osaki-ai}.  Substituting $q = p-1/2$, we get
\begin{equation}
I_{acc} = H\left({\textstyle \frac{1}{2}+q}\right) 
- H\left({\textstyle \frac{1}{2} + \frac{1}{2}\sqrt{1-\kappa (1-4q^2)} }
\right) 
\label{eq-accif}
\end{equation}
We wish to show that the second derivative of this quantity
is negative with respect
to $q$, for $-\frac{1}{2} < q < \frac{1}{2}$.  Let 
\[
R = 1-\kappa + 4\kappa q^2,
\]
which is the quantity under the radical sign in Eq.~(\ref{eq-accif}).

We now differentiate $I_{acc}$ with respect to $q$ and obtain
\begin{eqnarray*}
\frac{d^2I_{acc}}{dq^2} &=& H''\left({\textstyle \frac{1}{2}+q}\right) 
- \frac{4q^2\kappa^2}{R} H''\left({\textstyle
\frac{1}{2}+\frac{1}{2}\sqrt{R}}\right) 
- \frac{2\kappa(1-\kappa)}{R^{3/2}} 
H'\left({\textstyle\frac{1}{2}+\frac{1}{2}\sqrt{R}}\right) \\
&=&
-\frac{4}{1-4q^2} 
+ \frac{4q^2\kappa^2}{R} \frac{4}{\kappa(1-4q^2)} 
- \frac{2\kappa(1-\kappa)}{R^{3/2}} 
\ln \left(\frac{1-\sqrt{R}}{1+\sqrt{R}}\right)
\\
&=& 
\frac{2(1-\kappa)}{(1-4q^2) R^{3/2}} \left[ - 2 R^{1/2}  
+ \kappa (1-4q^2)
\ln\left(\frac{1+\sqrt{R}}{1-\sqrt{R}}\right)\right],
\end{eqnarray*}
which quantity we wish to show is negative.

We thus need to show that 
\[
\kappa (1-4q^2) \ln\left(\frac{1+\sqrt{R}}{1-\sqrt{R}}\right) \leq 2R^{1/2}
\]
Since $\kappa(1-4q^2) = 1-R$, this is equivalent to
\[
\log\left(\frac{1+\sqrt{R}}{1-\sqrt{R}}\right) \leq \frac{2\sqrt{R}}{1-R}.
\]
However, this is the inequality~(\ref{eq-inequality}) proven above, 
with $x=\sqrt{R}$, so we are done.

\vspace*{-.1in}

\section*{Appendix C: Accessible information for various $\alpha$.}
\begin{figure}[bth]
\begin{center}
\epsfig{figure=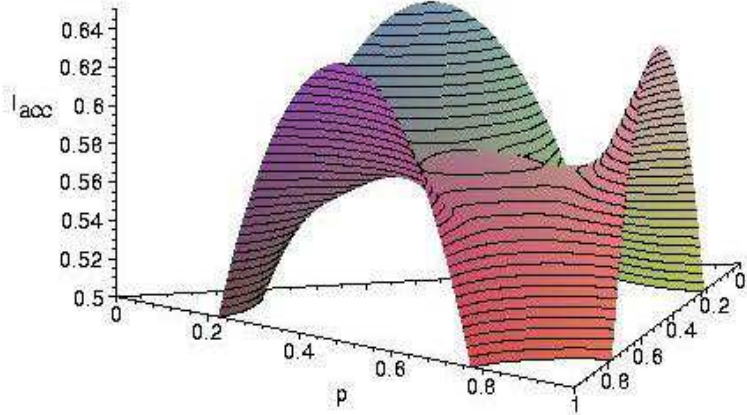,angle=270,width=4in}
\end{center}
\caption{\small 
The accessible information for 
the planar trines.  The probability distributions are represented
by a triangle where the vertexes correspond to probability distributions
$(1,0,0)$, $(0,1,0)$ and $(0,0,1)$.  
There are four local maxima, three at probability distributions symmetric
with $(0,\frac{1}{2},\frac{1}{2})$, and one at 
$(\frac{1}{3},\frac{1}{3},\frac{1}{3})$.
This was computed using a linear program,
considering as possible POVM elements the projectors 
$(\cos\theta,\sin\theta)$, for 36,000 evenly spaced values of $\theta$.  
The linear programming package CPLEX was used to calculate the
optimum for all probability distributions of the form 
$(\frac{a}{90}, \frac{b}{90}, \frac{c}{90})$, and this graph was drawn
by interpolating from these values.  We estimate the error for
each of these points $(\frac{a}{90}, \frac{b}{90}, \frac{c}{90})$
to be less than $10^{-5}$.  
\label{planartrines3d}
}
\end{figure}

In this section, we give graphs of the accessible information for
various values of $\alpha$.  These should be compared with Figure 
\ref{planartrines3d}, which gives the graph for the planar trines,
with $\alpha=0$.  This illustrates the origin of the behavior of 
the two crossing curves giving the $C_{1,1}$ capacity in 
Figure~\ref{fig-c1}.  The line BZ gives the value of the local 
maximum at the central point $(\frac{1}{3},\frac{1}{3},\frac{1}{3})$,
while the curve CY gives the behavior of the three at 
$(p, (1-p)/2, (1-p)/2)$.  It appears from numerical experiments that
this local maximum is achieved (or nearly achieved) using only three
projectors for $\alpha \leq 0.27$.  At a value of $\alpha$ slightly above 
$0.27$, the assumption that this local maximum is attained using 
a von Neumann measurement becomes false, and the curve of 
Figure~\ref{fig-c1}, which appears to give a local maximum of the 
information attainable using von Neumann measurements, no longer 
corresponds to a local maximum of the accessible information.
Note also that at the value $\alpha=0.27$, a POVM containing 
six projectors is required to achieve the $C_{1,1}$ capacity, even
though there are only three 3-dimensional states in the ensemble. 

\vspace*{-.2in}

\phantom{here}
\begin{figure}[htb]
\vspace*{-10pt}
\begin{center}
\epsfig{figure=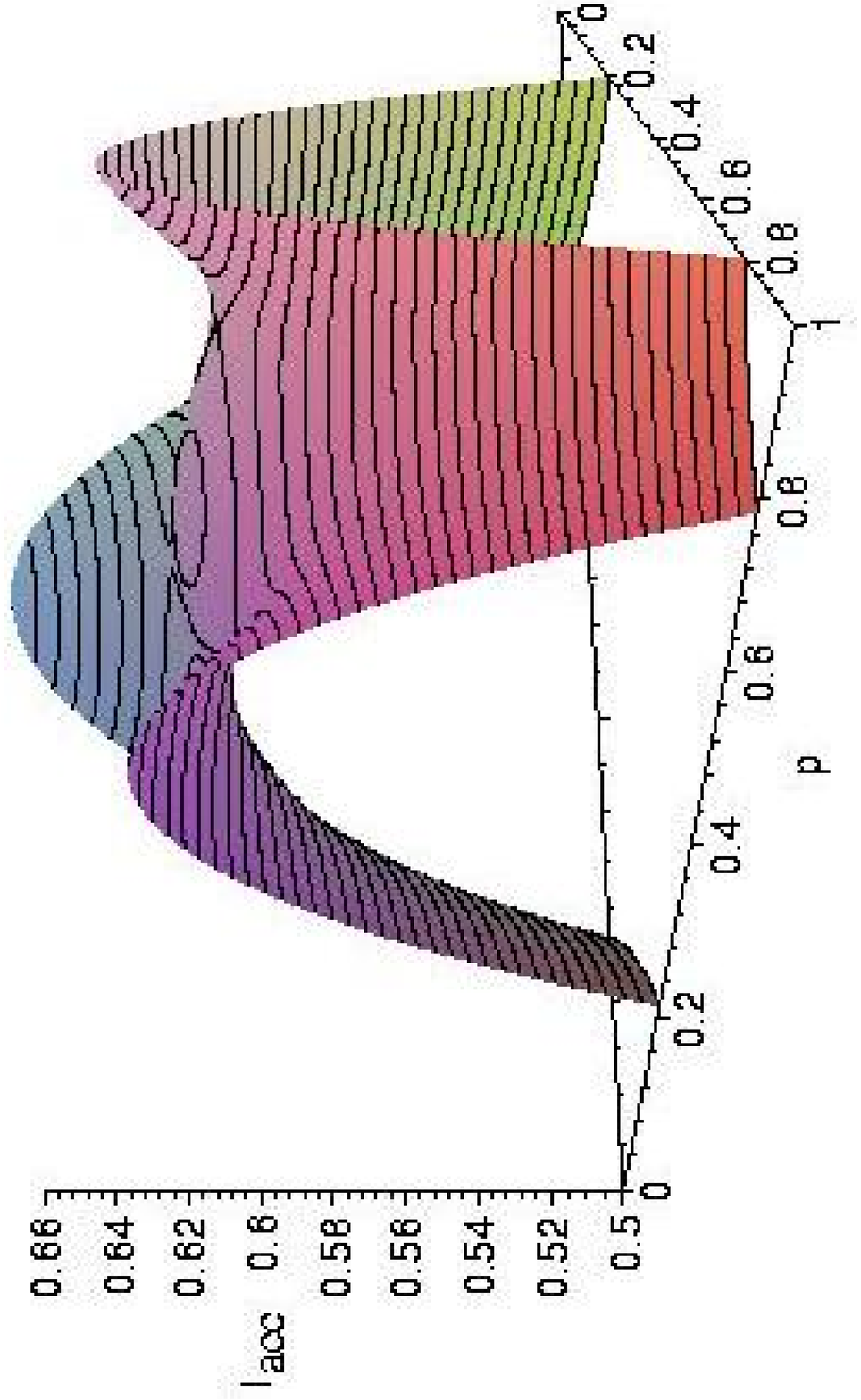,angle=270,width=4in}
\end{center}
\caption{\small 
The accessible information for
the trines with $\alpha=0.009$;  as in Fig.~\ref{planartrines3d}, 
the probability distributions are represented
by a triangle where the vertexes correspond to probability distributions
$(1,0,0)$, $(0,1,0)$ and $(0,0,1)$.  The maximum at $(0,0.5,0.5)$ (for
the planar trines) has moved slightly away from the edge; 
the maximum value now occurs
roughly at $(0.0012,0.4994,0.4994)$. The local maximum at 
$(\frac{1}{3},\frac{1}{3},\frac{1}{3})$ is growing larger with respect
to the global maximum.  This, and figures \protect{\ref{alpha.18}} and 
\protect{\ref{alpha.27}} were computed using the linear programming
package CPLEX to determine the optimal measurement for points of the form
$(\frac{a}{90}, \frac{b}{90}, \frac{c}{90})$, using projectors chosen
from 96,000 vectors distributed around the unit sphere.
\label{alpha.09}
}
\vspace*{-10pt}
\begin{center}
\epsfig{figure=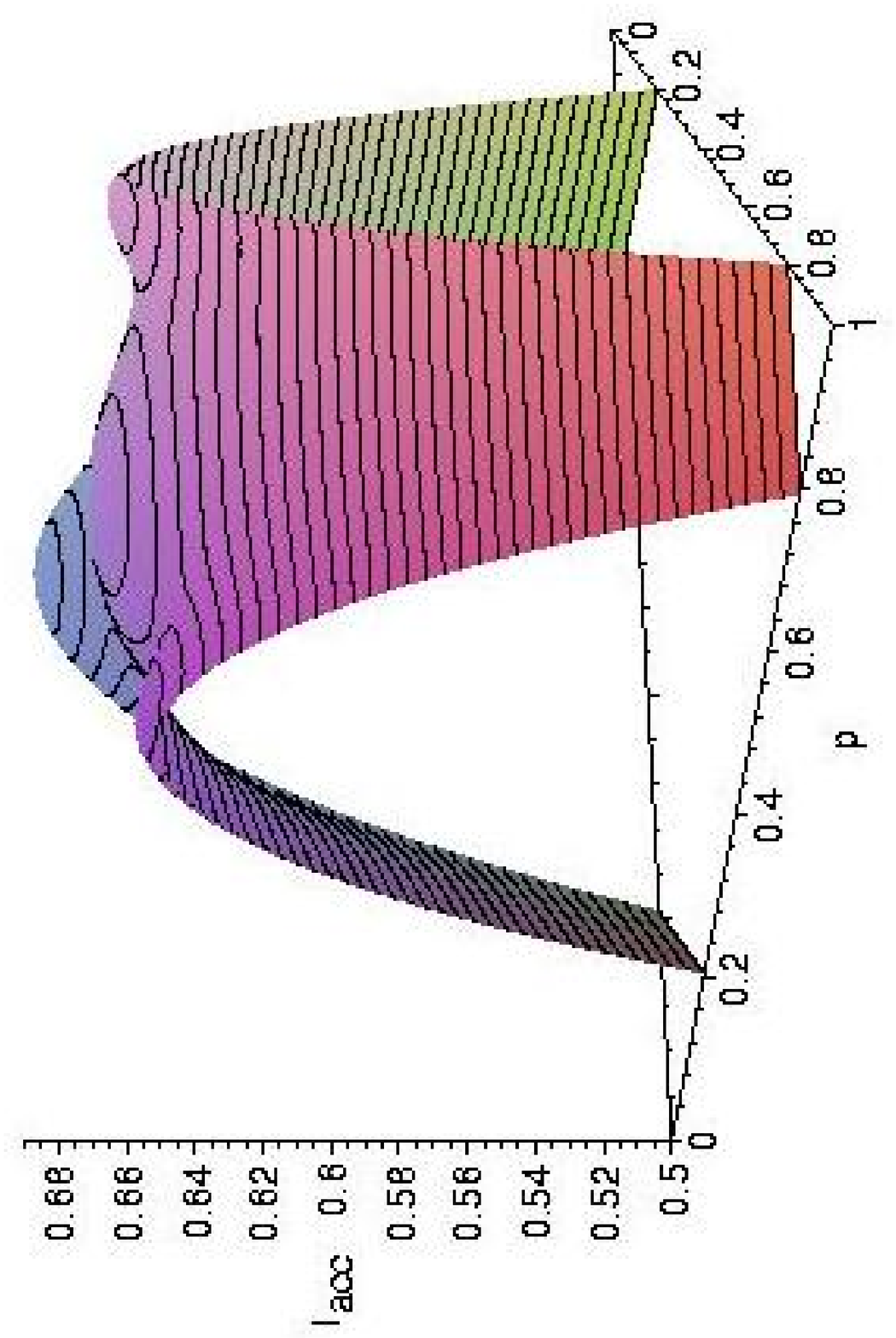,angle=270,width=4in}\phantom{tb}
\end{center}
\caption{\small 
The accessible information for
the trines with $\alpha=0.018$;  as in Fig.~\ref{planartrines3d}, 
the probability distributions are represented
by a triangle where the vertexes correspond to probability distributions
$(1,0,0)$, $(0,1,0)$ and $(0,0,1)$.   Here, the three local maxima have moved
farther in from the edges, and now occur at 
points symmetric with 
$(p,\frac{1-p}{2}, \frac{1-p}{2})$ for
$p\approx 0.027$; all four local maxima are now
nearly equal in value. 
\label{alpha.18}
}
\end{figure}
\begin{figure}[tbh]\phantom{tbh}
\vspace*{1in}
\begin{center}
\epsfig{figure=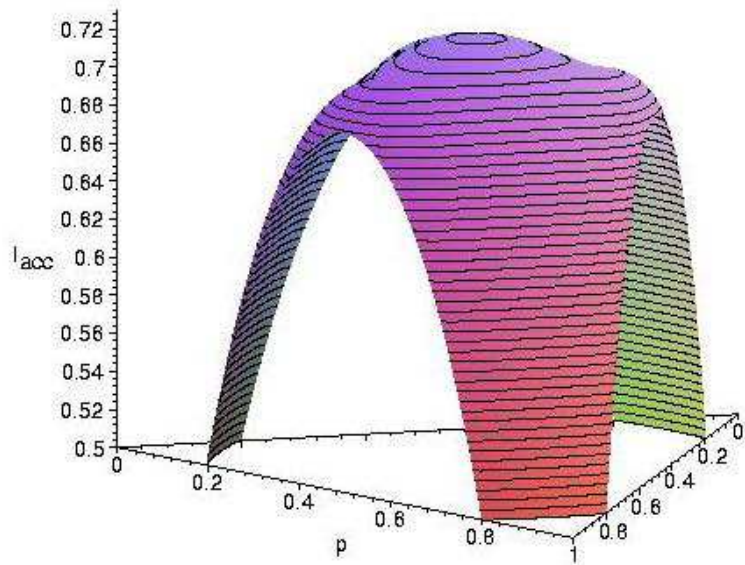,angle=270,width=4in}
\end{center}
\caption{\small 
The accessible information for
the trines with $\alpha=0.027$;  as in Fig.~\ref{planartrines3d}, 
the probability distributions are represented
by a triangle where the vertexes correspond to probability distributions
$(1,0,0)$, $(0,1,0)$ and $(0,0,1)$.  There are still four local maxima,
where the ones on the shoulders occur at points symmetric with 
$(p,\frac{1-p}{2}, \frac{1-p}{2})$ for $p \approx 0.105$;
these will disappear before $\alpha$ reaches $0.0275$.
\label{alpha.27}
}
\end{figure}

\end{document}